\documentclass[aps,prb,reprint,superscriptaddress,floatfix]{revtex4-2}
\usepackage[utf8]{inputenc}
\usepackage{hyperref}           
\usepackage{amsmath}
\usepackage{marvosym}
\usepackage{amssymb,stmaryrd}   
\usepackage{bbm}
\usepackage{babel}
\usepackage{comment}
\usepackage{bm}
\usepackage{color}
\usepackage{natbib}
\usepackage{graphicx,subfigure} 
\usepackage{epstopdf}
\usepackage{float}
\graphicspath{ {./figures/} }
\usepackage{dcolumn}    
\usepackage{wrapfig}
\usepackage{mathtools}
\usepackage{xspace}
\usepackage{refstyle}   
\usepackage{tikz}  
\usetikzlibrary{arrows,shapes,trees,positioning}  
\usepackage{siunitx}

\usepackage{appendix}
\usepackage{textcomp}
\usepackage{booktabs}
\usepackage[mathlines]{lineno}

\usepackage[normalem]{ulem}
\usepackage[ignoreunlbld,norefs,nocites]{refcheck}
\usepackage{multirow}
\hypersetup{
	colorlinks=true,
	linkcolor=blue,
	citecolor=blue,
	urlcolor=blue,
}

\definecolor{darkgreen}{rgb}{0, 0.8, 0}

\newcommand{\SNBNCBS}{\affiliation{Department of Condensed Matter and Materials Physics,
S. N. Bose National Centre for Basic Sciences, Block JD, Sector III, Salt Lake, Kolkata, 700106, India}}

\newcommand{\IOP}{\affiliation{Centre for Advanced Laser Techniques, Institute of Physics, Bijeni\v{c}ka Cesta 46, HR-10000 Zagreb, Croatia.}}

\newcommand{\Uppsalaphy}{\affiliation{Department of Physics and Astronomy, Uppsala University, Box 524, SE-751 20 Uppsala, Sweden}}

\newcommand{\austria}{\affiliation{Chair of Physics, Montanuniversit\"at Leoben, Franz Josef Strasse 18, 8700 Leoben, Austria}}

\newcommand{\Rijeka}{\affiliation{Faculty of Physics and Centre for Micro and Nanosciences and Technologies, University of Rijeka,
51000 Rijeka, Croatia}}

\begin{document}
\title{Observation of correlation-driven topological transport and robust ferromagnetism in 2D CrS$_2$}

\author{Sk Md Obaidulla}\email{Contact author: obaidulla20@gmail.com}

\IOP
\SNBNCBS

\author{M. Nur Hasan}
\Uppsalaphy

\author{Dayal Das}
\SNBNCBS

\author{Rafiqul Alam}
\SNBNCBS

\author{Antonio Supina}
\IOP

\author{Muhammad Awais Aslam}
\austria

\author{Sherif Kamal}
\IOP

\author{Iva \v{S}ari\'{c} Jankovic}
\Rijeka

\author{Aleksandar Matkovic}
\austria

\author{Christian Teichert}
\austria

\author{Atindra Nath Pal}
\SNBNCBS

\author{Heike C. Herper}
\Uppsalaphy

\author{Marko Kralj}
\IOP

\begin{abstract}
The realization of correlated layered magnets hosting robust ferromagnetism with emergent topological transport remains a key challenge in quantum materials. Here we report the first catalyst-free chemical vapour deposition growth of layered 1T-CrS$_2$, establishing a highly stable vdWs ferromagnet with an out-of-plane easy-axis anisotropy and a Curie temperature above room temperature. Transport measurements reveal a semimetal--insulator crossover near 80 K and pronounced negative magnetoresistance up to 350 K. A topological Hall effect emerges below 30 K, a rare signature of correlated transport in layered transition-metal dichalcogenide ferromagnets. First-principles calculations show that spin--orbit coupling gaps Dirac-like crossings, while electronic correlations reconstruct the Fermi surface by 
suppressing electron pockets and reducing the carrier density, enhancing momentum-dependent out-of-plane spin polarization. Magnetic measurements, supported by Heisenberg exchange calculations, reveal strong nearest-neighbour ferromagnetic exchange that stabilizes long-range ferromagnetism. Our results establish 1T-CrS$_2$ as a rare correlated 3$d$ layered ferromagnet in which electronic correlations and spin--orbit coupling cooperatively drive emergent topological transport.

\end{abstract}
\maketitle

\section{Introduction}

Two-dimensional (2D) van der Waals (vdW) magnets have emerged as an exciting platform for exploring correlated quantum phenomena arising from the interplay of magnetism, reduced dimensionality, and electronic topology \cite{bernevig2022progress,tokura2019magnetic,khan2020recent}. Although the Mermin–Wagner theorem predicts that thermal fluctuations suppress long-range magnetic order in isotropic two-dimensional (2D) systems at finite temperatures \cite{mermin1966absence}, finite magnetic anisotropy and strong exchange interactions can stabilize ferromagnetism even in the monolayer \cite{huang2017layer} and bilayer limits \cite{gong2017discovery}. At the same time, the interplay between intrinsic magnetism, electron correlations, and band topology gives rise to emergent quantum phenomena, including unconventional Hall responses and topologically nontrivial spin textures \cite{burkov2016topological}. Despite significant progress, the experimental realization of a single 2D platform that integrates these competing interactions in a robust and controllable manner remains limited. \par Cr-based 2D transition metal dichalcogenides (TMDCs), such as CrSe$_2$~\cite{li2021van}, and CrTe$_2$ \cite{tian2026room}, represent a distinct class of materials in which localized 3$d$ electrons give rise to strong correlation effects and intrinsic magnetism. The 1T polymorph, featuring octahedral coordination of the transition-metal atoms, hosts a rich variety of correlated and quantum phases, including the topological electronic phase reported in 1$T$$^\prime$-WTe$_2$ \cite{peng2017observation,fei2017edge} and the charge-density-wave state in 1T-TaS$_2$ \cite{alam2025nonlinear}. Unlike conventional TMDCs dominated by Fermi-surface instabilities, these systems exhibit correlation-driven electronic states that are highly sensitive to symmetry breaking and magnetic ordering. While recent studies suggest that such correlated magnetic backgrounds may host nontrivial Berry curvature and unconventional transport responses, experimentally establishing the interplay between band structure, magnetism, and transport remains a significant challenge \cite{xu2018electrically, adak2024tunable}. Semiconducting 2D Cr-based ferromagnets such as CrI$_3$ \cite{huang2017layer} and Cr$_2$Ge$_2$Te$_6$ \cite{gong2017discovery, gong2017discovery} require encapsulation due to environmental instability and exhibit limited electrical tunability, and often display reduced magnetic ordering temperatures or complex magnetic phase behavior. More importantly, the simultaneous realization of robust ferromagnetism, ambient stability, and tunable electronic topology within a single material system has remained elusive. The 1T polymorph of CrS$_2$ is particularly promising, but its phase-pure synthesis remains challenging owing to the thermodynamic stability of the competing 2H and 3R polytypes \cite{SafeerAFM, acsami.5c15768, bai2025phaseengineering1t1t}. Its octahedral crystal-field environment and edge-sharing CrS$_6$ octahedra give rise to a partially filled $t_{2g}$ ($d_{xy}$, $d_{yz}$, $d_{xz}$) manifold that promotes strong electron correlations, while the nearly 90$^\circ$ Cr--S--Cr bonding geometry naturally favors ferromagnetic superexchange interactions \cite{PhysRevB.97.245409}. Importantly, although spin–orbit coupling (SOC) is moderate in 3$d$ systems, its interplay with magnetic order can lift band degeneracies and reconstruct the low-energy electronic structure, providing a route to emergent transport phenomena. These features make 1T-CrS$_2$ an ideal system to explore how correlation-driven magnetism and SOC cooperate to shape electronic and transport properties in a layered 2D system.
\par Here we report a combined experimental and first-principles investigation of correlated ferromagnetism and unconventional electronic transport in highly stable, catalyst-free CVD-synthesized layered 1T-CrS$_2$. Magnetization measurements reveal robust ferromagnetic order persisting up to 350~K with an out-of-plane easy-axis magnetic anisotropy. Transport measurements further uncover a semimetal--insulator crossover near 80~K, persistent negative magnetoresistance, and the emergence of a topological Hall effect below 30~K, revealing a correlated transport regime. First-principles calculations show that SOC lifts band degeneracies near the Fermi level, reconstructing the electronic structure into a multiband state with strongly spin-polarized carriers, while strong nearest-neighbor Heisenberg exchange interactions stabilize the ferromagnetic ground state. Finite-temperature spin-dynamics simulations further predict a Curie temperature above 350~K, consistent with the experimental observations. Together, these findings establish 1T-CrS$_2$ as a rare correlated 3$d$ layered ferromagnet with intertwined electronic, magnetic, and topological transport properties, providing a promising platform for topological spintronics in two-dimensional quantum materials.

\section{Results and Discussions}

\begin{figure*}[t]
\centering
\includegraphics[width=\textwidth]{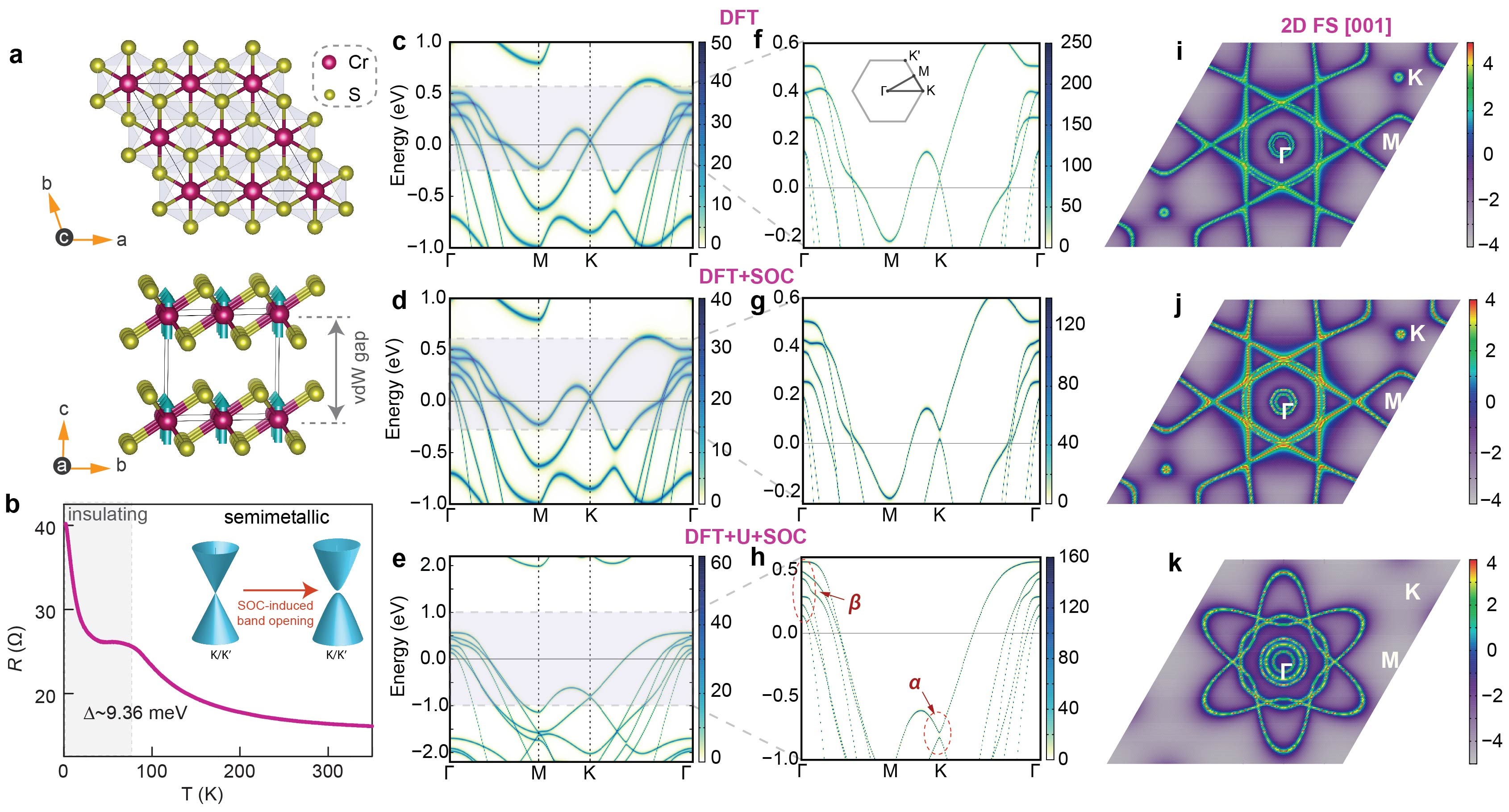}
\caption{\textbf{Crystal structure, electronic transport, spectral function, and 2D FS of 1T-CrS$_2$.} \textbf{a,} Top and side views of the 1T-CrS$_2$ crystal structure, illustrating the vdW gap between adjacent layers. Cr and S atoms are shown as cherry and yellow spheres, respectively. \textbf{b,} $R(T)$, showing a semimetal-to-insulator crossover. The inset schematically illustrates the lifting of the band degeneracy at the K/K$'$ points. \textbf{c--h,} Calculated spectral function (SF) along the high-symmetry path ($\Gamma$--M--K--$\Gamma$), indicated in the inset of \textbf{f}, obtained using \textbf{c,f} spin-polarized DFT without SOC, \textbf{d,g} DFT+SOC, and \textbf{e,h} DFT+$U$+SOC. In the absence of SOC (\textbf{c,f}), multiple dispersive bands cross the Fermi level. Including SOC (\textbf{d,g}) lifts the band degeneracy and induces band splitting at the Fermi level, while the inclusion of electronic correlations (\textbf{h}) suppresses the electron bands and opens SOC-induced gaps, labelled $\alpha$ at the K point and $\beta$ at the $\Gamma$ point. The colour scale represents the spectral intensity, with brighter colours indicating higher spectral weight. \textbf{i--k,} Calculated 2D FS projected onto the [001] plane, obtained using \textbf{i} spin-polarized DFT without SOC, \textbf{j} DFT+SOC, and \textbf{k} DFT+$U$+SOC. The adjacent colour scales represent the orbital contribution weight.}
\vspace{-0.45cm}
\label{fig:figure1}
\end{figure*}

\subsection{Semimetal–insulator crossover in a correlated 1T-CrS$_2$}

1T-CrS$_2$ crystallizes in a hexagonal lattice with space group $P\bar{3}m1$ (No.~164).  The crystal structure is composed of layered CrS$_6$ octahedra stacked along the $c$-axis  in the characteristic 1T coordination. Each Cr atom is coordinated by six S atoms,  forming octahedral CrS$_6$. These octahedra are edge-sharing within the $ab$ plane,  giving rise to a 2D layered framework. As shown in Fig. \ref{fig:figure1}a, the atomic positions are defined as follows: Cr atoms occupy the Wyckoff position $1a$ $(0,0,0)$,  while S atoms reside at $2d$ $(1/3,2/3,1/4)$ and $(1/3,2/3,3/4)$, respectively. In 1T-CrS$_2$, the Cr atoms form a triangular lattice within the crystallographic $ab$ plane, while the layers stack along the $c$-axis. Each individual layer consists of a trilayer S--Cr--S unit, where the S atoms are symmetrically positioned above and below the Cr plane within edge-sharing CrS$_6$ octahedra. Adjacent S--Cr--S layers are separated by an interlayer vdW gap, giving rise to the 2D crystal structure. 

\par Fig.~\ref{fig:figure1}b displays the temperature evolution of the longitudinal resistance $R(T)$  features the transport response of 1T-CrS$_2$. It reveals a pronounced crossover from semimetallic behavior at high temperatures to an insulating phase at low temperatures, with a characteristic transition temperature T$_{T_I}$ $\approx $80 K. Above this temperature, the resistance varies weakly with temperature and follows semimetallic transport dominated by electron–phonon scattering within itinerant hybridized between Cr 3$d$ and S 3$p$ orbitals. Below $T_{\mathrm{TI}} \approx 80$ K, the longitudinal resistance increases rapidly, indicating the emergence of an insulating transport regime. The low-temperature transport response is well described by a thermally activated behavior, \cite{kittel2018introduction,kumar2020topological} following $\rho(T) \propto \exp\left(\frac{\Delta}{k_B T}\right)$, from which an activation energy gap of $\Delta \approx 9.36$ meV is extracted. This small activation energy indicates the presence of a narrow electronic gap or band splitting near the Fermi level and suggests that the Fermi energy, E$_F$ lies close to a reconstructed electronic feature.

To investigate the microscopic origin of the transport crossover, the electronic structure and Fermi-surface (FS) topology of 1T-CrS$_2$ are investigated within progressively correlated theoretical frameworks. The spin-polarized DFT spectral function (SF) without SOC, calculated along the high-symmetry path $\Gamma$--M--K--$\Gamma$, is shown in Fig.~\ref{fig:figure1}c and Supplementary Fig.~S1. It exhibits a semimetallic character with multiple Cr-$t_{2g}$ bands crossing the Fermi level ($E_F$), giving rise to both electron- and hole-like carriers. A linearly dispersing crossing appears at the K valley, forming a Dirac-like band-crossing~\cite{PhysRevB.83.245125}. A magnified view of the low-energy dispersion is presented in Fig.~\ref{fig:figure1}f, where the two bands intersect at a single Dirac node in the absence of SOC~\cite{PhysRevB.83.245125}. The comparatively narrow bandwidth of these $t_{2g}$ states points to enhance correlation effects associated with the edge-sharing CrS$_6$ octahedral environment. The orbital-projected SF further shows that the low-energy electronic structure is governed predominantly by a subset of Cr-$t_{2g}$ orbitals (Supplementary Figs. S2-S6). In particular, the states crossing $E_F$ originate mainly from strongly hybridized in-plane $d$ orbitals, whereas several other orbital channels contribute weakly in the vicinity of the $E_F$. This orbital-selective character indicates that the semimetallic state is intrinsically anisotropic in orbital space even before incorporating SOC and explicit Coulomb correlations.

The corresponding calculated 2D [001] FS (Fig.~\ref{fig:figure1}) consists of multiple interconnected electron and hole pockets distributed around the $\Gamma$, K, and M points, characteristic of a multiband semimetal with relatively high carrier density and only weak reconstruction of the low-energy electronic states. Including SOC produces a clear reconstruction of the low-energy electronic structure, despite leaving the overall semimetallic band topology largely intact (Fig.\ref{fig:figure1}d). The most pronounced changes occur near the K valley and along the $\Gamma$--M direction, where previously degenerate bands become spin split due to the combined action of exchange polarization and relativistic spin--orbit interactions. As highlighted in the enlarged low-energy view shown in Fig.~\ref{fig:figure1}g, the linear Dirac-like crossing evolves into a gapped massive dispersion with finite momentum-dependent band splittings emerging throughout the vicinity of the $E_F$. These SOC-induced modifications introduce additional low-energy scales into the electronic states and substantially alter the topology of the near-$E_F$ states.

The orbital-projected SF reveals that SOC not only lifts the band degeneracies but also enhances the hybridization between different Cr-$d$ orbital sectors (Supplementary Figs.~S7--S11). While the low-energy electronic structure continues to be governed primarily by Cr-$t_{2g}$ states, the spectral weight (SW) becomes increasingly redistributed in momentum space, particularly around the K and M valleys where the reconstructed bands exhibit strong orbital mixing. This behaviour reflects the development of anisotropic spin--orbital entanglement within the low-energy electronic states. The influence of SOC is also evident in the corresponding 2D [001] FS shown in Fig.~\ref{fig:figure1}j. Relative to the non-SOC case, the Fermi contours become moderately distorted and anisotropic owing to the lifting of valley and spin degeneracies. Nevertheless, the overall multiband semimetallic character remains preserved, indicating that SOC primarily reconstructs the low-energy electronic states rather than driving a complete topological transformation of the FS. 

The inclusion of on-site Coulomb interactions within the DFT+$U$+SOC framework leads to a pronounced reorganization of the low-energy electronic structure, as shown in Fig.~\ref{fig:figure1}e and Supplementary Fig. S13. In contrast to the comparatively dispersive semimetallic bands obtained within DFT and DFT+SOC, the correlated electronic spectrum exhibits band renormalization accompanied by a significant suppression of SW near $E_F$. The resulting low-energy states become increasingly anisotropic and weakly dispersive, reflecting the formation of a correlated low-carrier-density state. Further insight into this reconstruction is provided by the orbital-resolved SF (Supplementary Figs.~S14--S18), which reveals a strongly orbital-selective redistribution of spectral intensity. While several Cr-$d$ orbital channels are significantly suppressed, the residual low-energy SW remains dominated by hybridized $t_{2g}$-derived states concentrated primarily around the K and M valleys. This behaviour indicates that Coulomb correlations and SOC cooperatively generate a momentum-dependent spin--orbital electronic behaviour within the low-energy electronic states. 

The reconstructed dispersion near $E_F$ is highlighted in the zoomed view Fig.~\ref{fig:figure1}h. The symmetry-protected Dirac-like crossing observed in the absence of SOC. In presence of SOC, it evolves into multiple finite-gap states labelled $\alpha$, and $\beta$. The dominant gap $\alpha$ appearing at the K-valley with additional smaller gaps near the $\Gamma$ point further reconstructs the low-energy electronic topology and the electron pockets at the $K$ and $M$ points of the Brillouin zone are suppressed. A corresponding reconstruction is observed in the calculated 2D [001] FS shown in Fig.~\ref{fig:figure1}k. The extended multiband semimetallic contours obtained within DFT progressively suppressed into much smaller disconnected pockets centred at $\Gamma$, reflecting a significant reduction in both carrier densities. Such an evolution is consistent with the experimentally observed insulating transport crossover and enhanced low-temperature topological Hall response. 

The SOC-induced $\alpha$ gap can be described within an effective massive Dirac framework, $H(\mathbf{k})=\hbar v_F\left(k_x\sigma_x+k_y\sigma_y\right)+\frac{\Delta_\alpha}{2}\sigma_z$, where $\mathbf{k}=(k_x,k_y)$ is the 2D crystal momentum measured from the valley center, $v_F$ is the Fermi velocity, $\sigma_x$, $\sigma_y$, and $\sigma_z$ are the Pauli matrices and $\Delta_\alpha$ is the temperature-dependent gap. The first term describes massless Dirac fermions and the second term introduces a finite mass arising from SOC and exchange-induced symmetry breaking. The corresponding dispersion relation is given by Eq.~(\ref{eq:dirac}) \cite{kozii2019thermoelectric,hu2025multipocket} describing massive Dirac quasiparticles separated by an energy gap ($\Delta_\alpha$).
\begin{equation}
E_{\pm}(\mathbf{k})=\pm \sqrt{(\hbar v_F k)^2+\left(\frac{\Delta_\alpha}{2}\right)^2},
\label{eq:dirac}
\end{equation}
Within the ferromagnetic phase, the low-energy electronic gap in 1T-CrS$_2$ is governed by the cooperative interplay of electronic correlations, exchange polarization, and SOC. Since both DFT and DFT+SOC calculations remain semimetallic, SOC alone is insufficient to generate a global gap. A finite gap emerges only in DFT+U+SOC calculations, indicating that the insulating-like state is primarily correlation driven and further reinforced by exchange splitting in the ferromagnetic phase. To capture the continuous semimetal--insulator crossover observed experimentally, we describe the temperature-dependent gap,
\begin{equation} 
\Delta_{\alpha}(T)=
\frac{\Delta_{SOC}+\lambda\left[m(T)-m_c\right]}
{1+\exp\left[-k\left(m(T)-m_c\right)\right]}, 
\label{eq:gap_T}
\end{equation}
where $\Delta_{SOC}$ is the correlation-induced gap, $m(T)$ is the temperature-dependent magnetization, $m_c$ denotes the critical magnetization required for FS reconstruction, $\lambda$ characterizes the exchange-enhanced contribution to the gap, and $k$ controls the sharpness of the crossover.

At low temperatures ($T\ll T_{\mathrm{TI}}$), the magnetization approaches its saturation value, resulting in enhanced exchange splitting and substantial reconstruction of the Cr-derived $d_{xy}$, $d_{xz}$, and $d_{yz}$ states near $E_F$. This suppresses the residual electron and hole pockets, stabilizing an insulating-like transport regime. As the temperature increases, thermal fluctuations progressively reduce the exchange polarization, causing the reconstructed gap to diminish continuously. When the temperatures well above the $T_{\mathrm{TI}}$ (i.e, $T\gg T_{\mathrm{TI}}$), the gap becomes negligible and the system gradually recovers a compensated spin-polarized semimetallic state, while long-range ferromagnetic order remains intact. This phenomenological framework describes the continuous temperature-driven reconstruction of the electronic structure, naturally reconciling the semimetal--insulator crossover with robust high-temperature ferromagnetism.

\begin{figure*}[t]
\centering
\includegraphics[width=\textwidth]{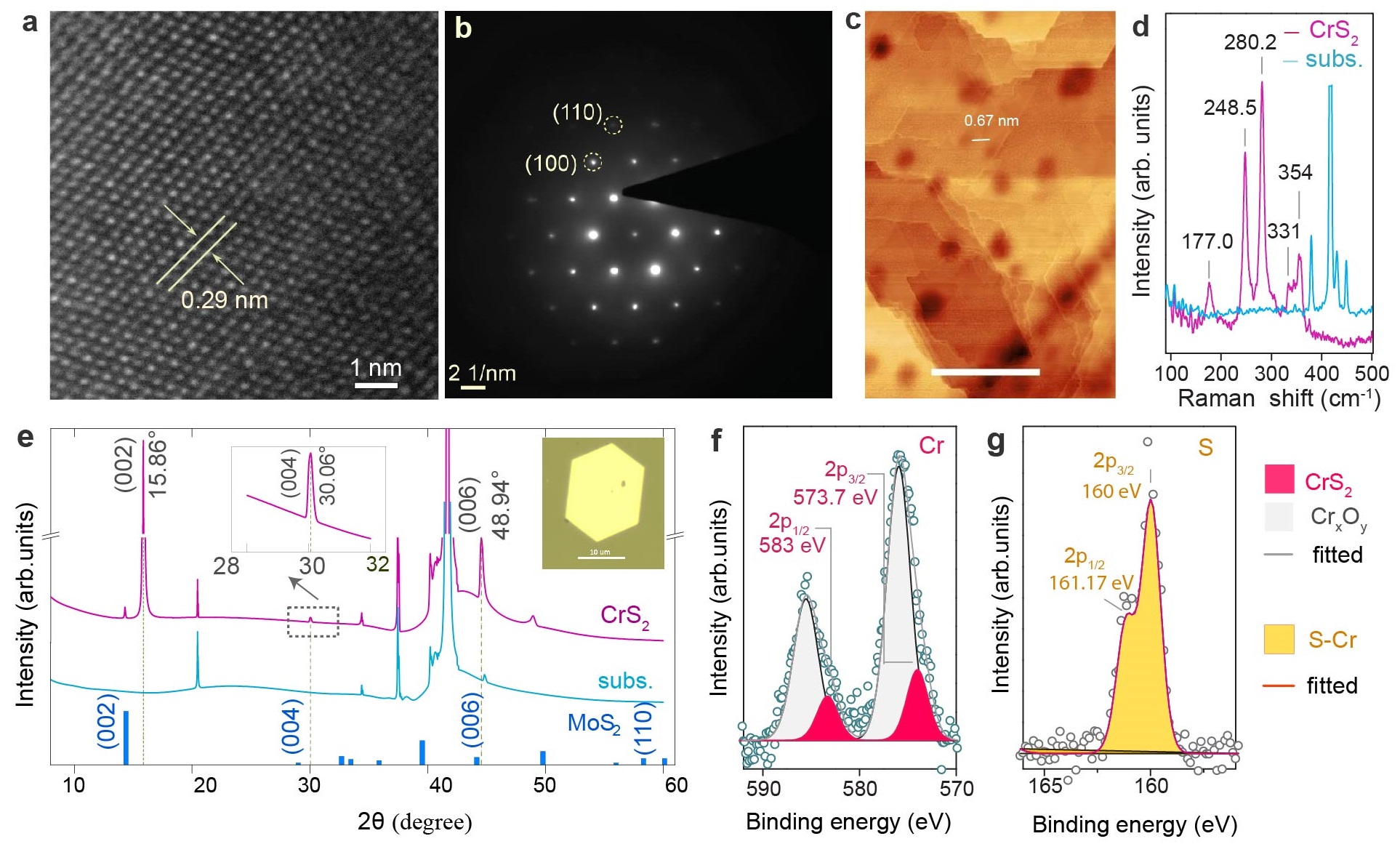}
\caption{\textbf{Structural and chemical characterization of 1T-CrS$_2$.}
\textbf{a,} HRTEM image of 1T-CrS$_2$, showing a well-ordered hexagonal lattice with an interplanar spacing of $\sim$0.29\,nm.
\textbf{b,} SAED pattern confirming the crystalline nature and hexagonal symmetry, indexed to the (100) and (110) planes.
\textbf{c,} AFM of CrS$_2$ with step-height (0.67 nm) displaying layered nature. Scale bar: 1 $\mu$m.
\textbf{d,} Raman spectra of CrS$_2$ (magenta) compared with background signal (cyan), displaying characteristic vibrational modes.
\textbf{e,} XRD pattern of CrS$_2$ (cherry) with background (cyan), and showing prominent (00l) reflections, consistent with a layered structure. Reference peaks from CVD-synthesized MoS$_2$ are included for comparison. Inset: zoomed-in view of (004) peak and hexagonal CrS$_2$ crystal.
\textbf{f,} XPS spectrum of the Cr 2p region, showing spin--orbit split peaks corresponding to Cr 2p$_{3/2}$ and Cr 2p$_{1/2}$, indicative of the Cr oxidation state in CrS$_2$.
\textbf{g,} XPS spectrum of the S 2p region with fitted components, revealing dominant S--Cr bonding, along with traces of surface oxide (Cr$_x$O$_y$). The binding energies are consistent with Cr$^{4+}$ and S$^{2-}$ species.}
\vspace{-0.45cm}
\label{fig:figure2}
\end{figure*}
\subsection{Catalyst-Free CVD Growth of Phase-Pure 1T-CrS$_2$}
 An optimized catalyst-free CVD process (Supplementary Figs.~S19-20) yields large ($>15\,\mu$m), well-faceted hexagonal and triangular single crystals suitable for exfoliation and device fabrication. The atomic structure of a 1T-CrS$_2$ monolayer, illustrated in Figs \ref{fig:figure1}a,b is composed of a single slab of edge-sharing CrS$_6$ octahedra. This octahedral coordination of the Cr-atom is the defining structural characteristic of the 1T-phase and is the fundamental origin of its unique electronic properties. In bulk, plate-like 1T-CrS$_2$ crystal consist of multiple such monolayers stacked along the c-axis $via$ vdW interactions. Thus, our successful synthesis of high-quality 1T-CrS$_2$ provides the essential material foundation for this investigation.


To provide direct real-space and reciprocal-space verification, we conducted transmission electron microscopy (TEM) and selected-area electron diffraction (SAED). The high-resolution TEM (HR-TEM) image (Figure~\ref{fig:figure2}a) resolves a defect-free hexagonal lattice with interplanar spacings of $\sim$0.29 and $\sim$0.18 nm, indexed to the (100) and (110) planes, respectively, consistent with the 1T crystal structure (Figure~\ref{fig:figure2}b). The lattice parameter extracted from the SAED analysis of 1T-CrS$_2$ using the hexagonal lattice formula (Supplementary Section~A) was found to be $a = 3.335$~\AA, determined from the (100) planes. This value is in excellent agreement with the theoretically optimized lattice parameter of 1T-CrS$_2$ ($a = 3.34$~\AA), confirming the formation of a phase-pure octahedral 1T structure. The SAED pattern (Fig.~\ref{fig:figure2}b) exhibits hexagonal symmetry and is indexed along the [001] zone axis, confirming the octahedral 1T structure and long-range crystalline order, consistent with single-crystal MoS$_2$ \cite{lee2012synthesis}. The presence of only one set of hexagonal diffraction patterns in the SAED image in Fig.~\ref{fig:figure2}b confirms the single-crystalline nature of the CVD-grown highly pure CrS$_2$ crystal. The absence of streaking, superlattice spots, or additional diffraction rings further confirms the high phase purity and crystallographic quality of our synthesized CrS$_2$. Atomic force microscopy (AFM) was used to characterize the morphology and thickness of the CVD-grown flakes. AFM imaging and height profiling (Fig.~\ref{fig:figure2}c) reveal atomically thin CrS$_2$ flakes with clean surfaces and sharp step edges, characteristic of high-quality van der Waals crystals. The measured step height of $\sim$0.7 nm is consistent with a single S–Cr–S layer, confirming the ultrathin layered nature of the as-grown 2D CrS$_2$ crystals and their potential for nanoscale device integration. The Raman spectrum of 1T-CrS$_2$ (Fig.~\ref{fig:figure2}d) exhibits several characteristic peaks, which are distinctly different from the 2H phase. The Raman spectrum displays three characteristic modes, R$_1$ (177.0~cm$^{-1}$), R$_2$ (248.5~cm$^{-1}$), and R$_3$ (280.2~cm$^{-1}$), consistent with the 1T structure of MoS$_2$ \cite{lei2018recent}. The R$_2$ and R$_3$ modes correspond to out-of-plane and in-plane vibrations, respectively, with a separation of $\Delta \approx 31.7$~cm$^{-1}$, matching the reported value of 32~cm$^{-1}$ \cite{shivayogimath2019universal}, confirming the octahedral coordination of 1T-CrS$_2$. The minor deviation is expected due to the use of a different laser source. However, it is noted that $\Delta \approx 33$--34~cm$^{-1}$ is reported for Cr$_2$S$_3$ crystals, which is higher than our observation for CrS$_2$ crystals \cite{cui2020controlled}. The narrow linewidths and strong Raman intensities further indicate the high crystalline nature of the as-grown CrS$_2$ flakes. XRD measurements provide further confirmation of the crystal structure and phase purity. The XRD pattern (Figure~\ref{fig:figure2}e) exhibits sharp (00$l$) reflections corresponding to the (002) and (004) planes at 15.86° and 30.06°, respectively, confirming the high crystallinity and vdWs stacking along the $c$-axis. The chemical composition and bonding environment were further examined by X-ray photoelectron spectroscopy (XPS) (Fig.~\ref{fig:figure2}f,g). Core-level spectra of Cr~2$p$ and S~2$p$ were acquired immediately after exfoliation and remeasured after one year of ambient exposure, revealing negligible changes in peak positions and spectral line shapes. The Cr~2$p$ spectra (Fig.~\ref{fig:figure2}f) are well described by two spin--orbit-split doublets with a characteristic separation of $\sim$9.5~eV between the Cr~2$p_{3/2}$ and Cr~2$p_{1/2}$. The dominant doublet, centered at $\sim$576~eV, is assigned to Cr$^{4+}$ bonded to sulfur in the 1T-CrS$_2$ lattice, consistent with reported binding energies for CrO$_2$ and clearly distinct from those of Cr$_2$S$_3$ \cite{wang2019electronic, rajendran2022chemical}. The S~2$p$ spectrum (Fig.~\ref{fig:figure2}g) is similarly fitted with two spin--orbit-split doublets corresponding to the S~2$p_{3/2}$ and S~2$p_{1/2}$. The lower-binding-energy doublet (160.0 and 161.0 eV) is assigned to lattice S$^{2-}$ bonded to Cr, whereas the weak higher-binding-energy component ($\sim$169 eV) originates from surface S--O species. The spectral features indicate minimal surface oxidation and a robust local electronic structure in 1T-CrS$_2$.

\begin{figure*}[t]
\centering
\includegraphics[width=\textwidth]{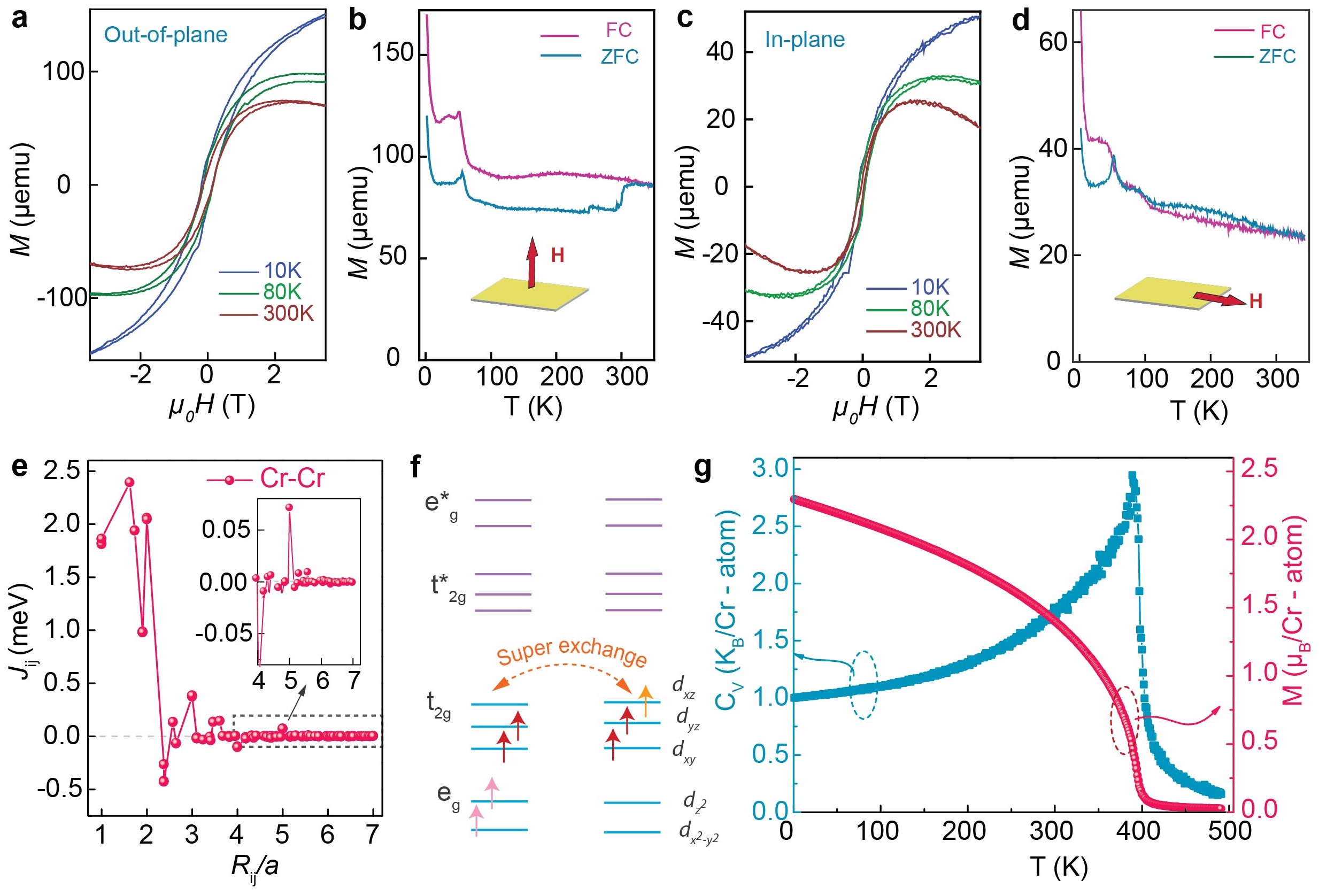}
\caption{\textbf{Magnetic properties, Heisenberg exchange interactions, and finite-temperature behavior of 1T-CrS$_2$.}
\textbf{a,b,} Out-of-plane field-dependent magnetization, $M(H)$, and temperature-dependent magnetization, $M(T)$. The hysteresis loop confirms an out-of-plane easy axis.
\textbf{c,d,} In-plane field-dependent magnetization, $M(H)$, and temperature-dependent magnetization, $M(T)$. Temperature-dependent magnetization \textbf{b,c,}, $M(T)$ measured under an applied field of 100~Oe.
\textbf{e,} Calculated Heisenberg exchange coupling parameters, $J_{ij}$, as a function of the normalized Cr--Cr distance ($R_{ij}/a$). The nearest-neighbor interaction is strongly ferromagnetic.
\textbf{f,} Schematic illustration of the superexchange mechanism in 1T-CrS$_2$.
\textbf{g,} Calculated $M(T)$, and $C_V$ using Monte Carlo simulations. A sharp peak in $C_V$ at $T_{\mathrm{C}}\sim390$~K marks the magnetic phase transition.}
\vspace{-0.45cm}
\label{fig:figure3}
\end{figure*}

\subsection{Room-temperature ferromagnetism and Heisenberg exchange interactions in 1T-CrS$_2$}

Fig.~\ref{fig:figure3} shows the magnetic properties of 1T-CrS$_2$ through a combination of experimental magnetization measurements and first-principles-based theoretical calculations. The field-dependent magnetization $M(H)$, measured with the magnetic field applied both out-of-plane ($H \perp ab$) and in-plane ($H \parallel ab$), as shown in Fig.~\ref{fig:figure3}a and \ref{fig:figure3}c, reveals clear ferromagnetic hysteresis at low temperatures. At 10~K, the out-of-plane magnetization reaches a saturation value of approximately $(1.3$--$1.4)\times10^{-4}$ emu under an applied field of $\sim$3~T, whereas the in-plane configuration saturates at a significantly lower value of $(4$--$5)\times10^{-5}$ emu even at comparable fields. The pronounced anisotropy between the two field orientations identifies the out-of-plane direction as the magnetic easy axis of 1T-CrS$_2$. The magnetic anisotropy energy (MAE) extracted from magnetization measurements is $0.28\,\mathrm{MJ\,m^{-3}}$ at 10~K, confirming the uniaxial magnetic anisotropy. The corresponding coercive field (Fig.~\ref{fig:figure3}a) remains relatively small (on the order of ~100–200 Oe), indicating soft ferromagnetic behaviour with low domain pinning \cite{ge2021direct}. The anisotropy persists up to higher temperatures (300 K), although both the saturation magnetization and the hysteresis progressively decrease, reflecting enhanced thermal spin fluctuations. The ratio of out-of-plane to in-plane magnetization remains larger ($\sim$ 10 times) across the measured temperature range, quantitatively confirming the robustness of anisotropic spin alignment. This behavior originates from the interplay between crystal field symmetry and spin–orbit coupling within the CrS$_6$ octahedral environment, which energetically stabilizes spins perpendicular to the layers. It is noted that Fig. \ref{fig:figure3}a and \ref{fig:figure3}c exhibit distinguishable and with enhanced magnetization without reaching saturation up to 3 $T$, suggesting typical semimetal behavior \cite{hu2025multipocket}. The temperature-dependent magnetization $M(T)$ as shown in Fig.~\ref{fig:figure3}b,d further supports robust ferromagnetism. For both out-of-plane and in-plane magnetic-field orientations, $M(T)$ increases sharply at low temperatures upon cooling, followed by a gradual decrease with increasing temperature. A clear bifurcation between field-cooled (FC) and zero-field-cooled (ZFC) curves is observed below approximately $T \sim 80\,\mathrm{K}$, indicating the onset of magnetic ordering and domain freezing effects. The finite magnetization retained up to room temperature ($\sim 300\,\mathrm{K}$) signifies the survival of strong ferromagnetic correlations well above the low-temperature regime. The enhanced FC--ZFC splitting in the out-of-plane geometry further supports the presence of appreciable magnetic anisotropy. Since no evident magnetic transition is observed up to 350\,K, the Curie temperature of 1T-CrS$_2$ is expected to exceed this value, establishing the system as a promising high-temperature 2D ferromagnet.

The magnetic anisotropy energy was calculated from first principles according to~\cite{PhysRevLett.75.2871}
\begin{equation}
\mathrm{MAE}=E_{\parallel ab}-E_{\perp ab},
\end{equation}
where $E_{\parallel ab}$ and $E_{\perp ab}$ denote the total energies for in-plane and out-of-plane magnetization orientations, respectively. The calculated MAE by DFT+$U$+SOC method is found to be $0.27\,\mathrm{MJ\,m^{-3}}$, in close agreement with the experimental value ($0.28\,\mathrm{MJ\,m^{-3}}$ at 10 K). In this respect, the finite magnetic anisotropy plays a crucial role in stabilizing the ferromagnetic ground state in accordance with the Mermin--Wagner theorem for low-dimensional magnetic systems~\cite{wang2020prospects}. The microscopic origin of the ferromagnetic ground state is further investigated by the magnetic exchange interactions calculated using the Liechtenstein--Katsnelson--Antropov--Gubanov (LKAG) formalism~\cite{LIECHTENSTEIN198765} implemented within the RSPt code~\cite{wills2000full}. Within this approach, the system is described by a generalized classical Heisenberg model,

\begin{equation}
H=-\sum_{i\neq j} J_{ij}\,\mathbf{e}_{i}\cdot\mathbf{e}_{j},
\label{eq:heisenberg}
\end{equation}

where $\mathbf{e}_{i}$ is the unit vector defining the orientation of the local magnetic moment at site $i$ and $J_{ij}$ denotes the isotropic exchange interaction between magnetic moments located at sites $i$ and $j$. In the nonrelativistic limit $J_{ij}$ is the change in the total electronic energy associated with infinitesimal rotations of two magnetic moments given by,

\begin{equation}
J_{ij}=\frac{1}{4\pi}\,\mathrm{Im}
\int_{-\infty}^{E_{\mathrm{F}}}
\mathrm{Tr}
\left[
\hat{\Delta}_{i}
\hat{G}^{\uparrow}_{ij}(\varepsilon)
\hat{\Delta}_{j}
\hat{G}^{\downarrow}_{ji}(\varepsilon)
\right]
d\varepsilon,
\label{eq:lkag}
\end{equation}

where $E_{\mathrm{F}}$ is the Fermi energy, $\hat{\Delta}_{i}$ is the onsite exchange-splitting potential, and $\hat{G}^{\uparrow,\downarrow}_{ij}$ are the spin-dependent intersite Green's functions. This formalism establishes a direct connection between the electronic structure and the magnetic exchange interactions, enabling an accurate first-principles description of magnetic coupling in itinerant magnetic systems.

Fig.~\ref{fig:figure3}e shows the calculated isotropic exchange parameters $J_{ij}$ as a function of the normalized Cr--Cr distance ($R_{ij}/a$). The calculated exchange interactions are predominantly ferromagnetic and remain finite over several neighbouring shells. Interestingly, although the nearest-neighbour exchange interaction is ferromagnetic, the second- and third-nearest-neighbour couplings become comparatively stronger. This behaviour reflects the itinerant nature of magnetic exchange in 1T-CrS$_2$, where indirect Cr--S--Cr hybridization enhances intermediate-range exchange pathways beyond the shortest Cr--Cr distance. Such an exchange hierarchy originates from the edge-sharing CrS$_6$ octahedral environment characteristic of the 1T structure. While the nearest-neighbour Cr--S--Cr bond geometry approaches $\sim90^\circ$, favouring ferromagnetic superexchange, the longer-range exchange channels benefit from enhanced orbital overlap mediated through hybridized Cr-$3d$ and S-$3p$ states. Beyond the nearest neighbours, the exchange interactions decay rapidly toward negligible values, indicating the absence of significant competing antiferromagnetic frustration.

\par The microscopic origin of this ferromagnetic coupling is illustrated schematically in Fig.~\ref{fig:figure3}f. In the octahedral crystal field of the 1T phase, the Cr $3d$ orbitals split into lower-energy $t_{2g}$ ($d_{xy}$, $d_{yz}$, $d_{xz}$) and higher-energy $e_g$ ($d_{x^2-y^2}$, $d_{z^2}$) states. For the $3d^{2}$ electronic configuration, the electrons occupy the $t_{2g}$ manifold, forming localized $S=1$ moments \cite{PhysRevB.97.245409}. The nearly 90$^\circ$ Cr--S--Cr bonding geometry favors ferromagnetic superexchange, in which virtual hopping through the S-$p$ orbitals stabilizes parallel spin alignment, consistent with the Goodenough--Kanamori rules~\cite{soriano2020magnetic}. The combined effects of ferromagnetic exchange, magnetic anisotropy, and electronic correlations stabilize the ferromagnetic ground state of 1T-CrS$_2$.

The finite-temperature magnetic behaviour was further investigated by Monte Carlo method performed within the UppASD package~\cite{Eriksson2017}. The calculated $M(T)$ and specific heat $C_V$ are shown in Fig.~\ref{fig:figure3}g. The saturation magnetization decreases continuously and vanishes at $T_{\mathrm{C}}=390$~K, coincident with a pronounced peak in $C_V$ that identifies a continuous second-order ferromagnetic phase transition. The calculated magnetic T$_C$, significantly exceeding room temperature, is consistent with the experimentally measured magnetization. Table~\ref{tab:Cr_materials}, 1T-CrS$_2$ distinguishes itself from representative Cr-based and other 2D TMDC magnets by combining a high T$_C$ with exceptional environmental stability. 

\begin{table*}[t]
\centering
\caption{Comparison of the Curie temperature, environmental stability, and Hall response of representative 2D TMDC magnets with 1T-CrS$_2$.}
\label{tab:Cr_materials}

\small
\renewcommand{\arraystretch}{1.15}
\setlength{\tabcolsep}{5pt}

\begin{tabular*}{\textwidth}{@{\extracolsep{\fill}} l l l c l l l}
\hline
\textbf{TMDC} & \textbf{Thickness} & \textbf{Hall response} & \textbf{$T_{\mathrm{C}}$ (K)} & \textbf{Stability} & \textbf{Method} & \textbf{Ref.} \\
\hline

1T-VSe$_2$     & 1L, 2L      & Anomalous$^a$   & $>300$   & --        & MBE$^b$ & \cite{bonilla2018strong} \\
1T$'$-MnS$_2$  & 0.92        & Anomalous       & $>360$   & --        & CVD     & \cite{zhao2025experimental} \\
CoS$_2$        & 0.92        & Anomalous       & 120--128 & --        & CVD     & \cite{zhao2025experimental} \\
CrI$_3$        & 4.2         & Anomalous       & 40       & --        & Exfoliation   & \cite{huang2017layer} \\
1T-CrTe$_2$    & 10          & Anomalous       & --       & 5 days    & CVD     & \cite{meng2021anomalous} \\
CrSe$_2$       & 4.1         & Anomalous       & --       & 45 days   & CVD     & \cite{li2021van} \\
1T-CrS$_2$     & Multilayer  & Topological$^c$ & $>350$   & $>1$ year & CVD     & This work \\

\hline
\end{tabular*}

\vspace{1mm}

{\footnotesize
$^a$ Anomalous Hall Effect \quad
$^b$ Molecular beam epitaxy \quad
$^c$ Topological Hall Effect
}

\end{table*}
\subsection{An emergent Topological Hall fingerprint and its microscopic origin}
\begin{figure}[t]
    \centering
    \includegraphics[width=\columnwidth]{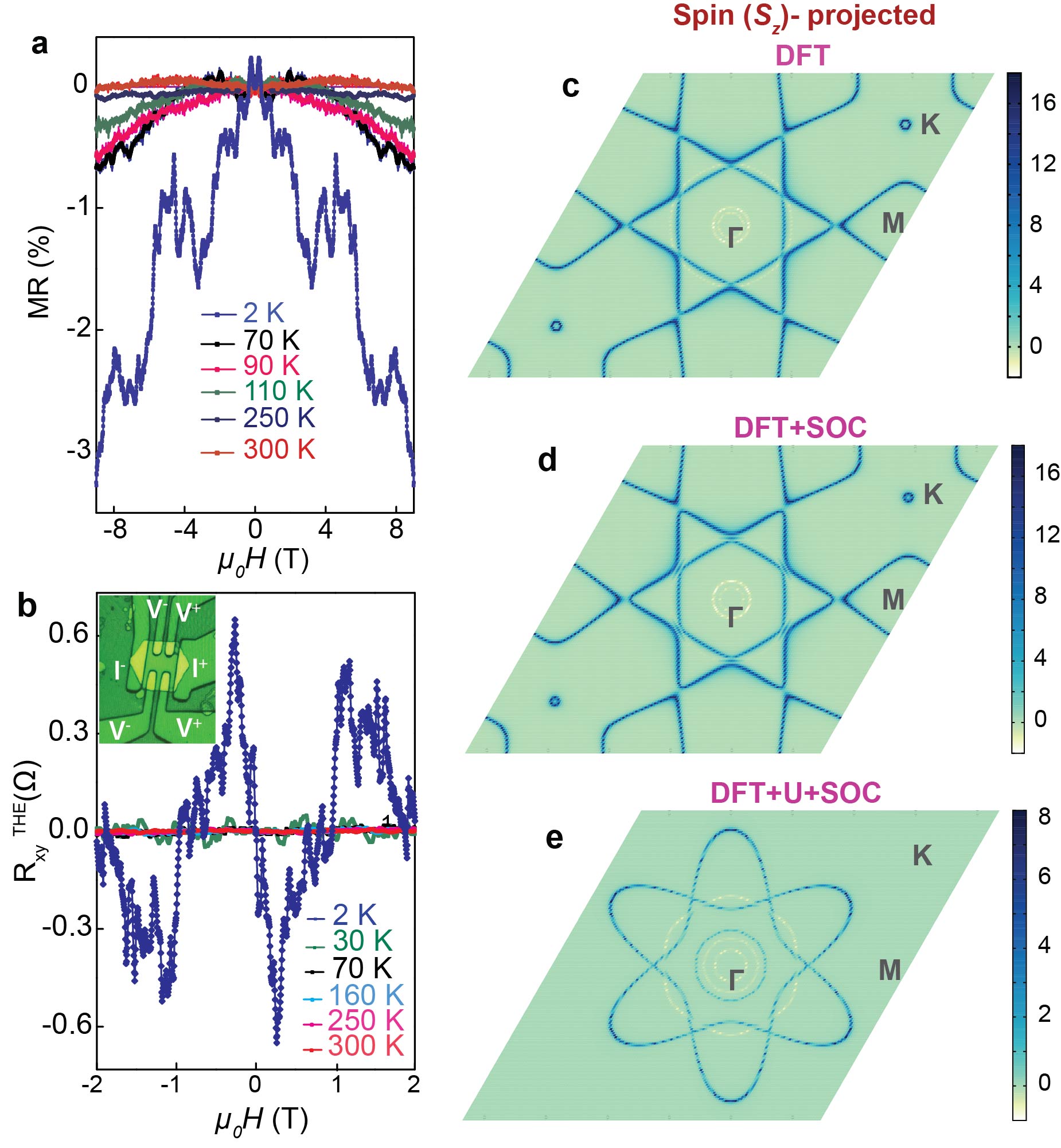}
    \caption{\textbf{Magnetotransport and partial FS topology of 1T-CrS$_2$.}
    \textbf{a,} MR measured at different temperatures (2--300\, K), showing a pronounced negative MR at low temperatures.
    \textbf{b,} Hall resistance $R^{THE}_{xy}$ as a function of magnetic field at various temperatures. The nonlinear behaviour at low temperatures indicates multiband transport, whereas the nearly linear response at higher temperatures suggests a dominant single-carrier contribution. Inset: Optical image of the device and the electrical-contact configuration used for transport measurements.
    \textbf{c--e,} Calculated 2D spin-projected ($S_z$) FS. The calculations were performed using spin-polarized DFT without SOC (\textbf{c}), fully relativistic DFT+SOC (\textbf{d}), and DFT+$U$+SOC (\textbf{e}). Colour scales represent the variation in carrier dynamics across different regions of the FS.}
    \label{fig:figure4}
\end{figure}

Figure~\ref{fig:figure4}a presents the longitudinal magnetoresistance, defined as $\mathrm{MR}=[R_{xx}(B)-R_{xx}(0)]/R_{xx}(0)$~\cite{jiang2017skyrmions,roychowdhury2024giant}, measured for $H\perp ab$ over 2--300~K.
The MR is predominantly negative throughout the entire temperature range, with no appreciable conventional positive orbital contribution even at 300~K. Its magnitude is largest at 2~K, reaching approximately $-3.2\%$ at 8~T, and decreases progressively with increasing temperature to below $0.1\%$ at 300~K. The persistence of negative MR up to room temperature is unusual for layered TMDCs~\cite{jie2017observation}, where Lorentz-force-driven orbital scattering typically gives rise to negligible positive MR under out-of-plane magnetic fields~\cite{ali2014large}. The observed behavior instead indicates that magnetotransport in 1T-CrS$_2$ is dominated by spin-dependent scattering. Below $T_{\mathrm{C}}$ ($350$~K), the applied magnetic field suppresses thermally driven spin fluctuations, thereby reducing spin-disorder scattering and the electrical resistivity. The weak orbital contribution suggests that charge transport remains strongly coupled to the magnetic configuration throughout the ferromagnetic state. This coupling is further enhanced by the layered vdWs structure of 1T-CrS$_2$, where weak interlayer hybridization and anisotropic magnetic interactions increase the sensitivity of carrier transport to the underlying spin texture~\cite{soriano2020magnetic}. Consequently, enhanced thermal spin disorder progressively weakens the field-induced reduction of spin-disorder scattering, accounting for the observed decrease in the magnitude of the negative MR with increasing temperature.

To extract the topological Hall contribution, the antisymmetrized Hall resistance measured during increasing ($-9 \rightarrow 9$~T) and decreasing ($9 \rightarrow -9$~T) magnetic-field sweeps. The topological Hall resistance is extracted as $R_{xy}^{\mathrm{THE}}=\Delta R_H/2=[R_{xy}^{\uparrow}(B)-R_{xy}^{\downarrow}(B)]/2$, where $R_{xy}^{\uparrow}$ and $R_{xy}^{\downarrow}$ denote the antisymmetrized Hall resistances recorded during the increasing- and decreasing-field sweeps, respectively \cite{piva2023topological}. This procedure effectively isolates the hysteretic topological Hall contribution while suppressing the reversible ordinary and anomalous Hall components. The residual Hall signal, $R_{xy}^{\mathrm{THE}}$ (Fig.~\ref{fig:figure4}b), exhibits a pronounced hump-like feature at intermediate magnetic fields ($\sim$0--2~T), revealing the emergence of a topological Hall effect (THE). The THE is strongest at 2~K and gradually weakens with increasing temperature, vanishing above $\sim$30~K. The confinement of the THE response to the low-temperature regime indicates that it is not simply associated with the high-temperature ferromagnetic state alone, but instead reflects a temperature-dependent reconstruction of the electronic and spin-polarized environment in layered 1T-CrS$_2$ system. The emergence of the THE below $\sim 30$~K is particularly significant in the context of the stable ferromagnetic phase of 1T-CrS$_2$, as in conventional ferromagnets, the Hall response is typically dominated by ordinary and anomalous Hall contributions alone \cite{wang2022topological, huang2021two}. The appearance of an additional low-temperature topological Hall component therefore signals the onset of a distinct transport regime in which charge carriers experience an additional transverse scattering process beyond conventional ferromagnetic transport. This interpretation is further supported by the simultaneous enhancement of the negative MR and the emergence of insulating-like transport behaviour below $\sim 80\,\mathrm{K}$.

To elucidate the microscopic origin of the observed transport behavior, the spin-polarized ($S_z$) FS are shown in Fig.~\ref{fig:figure4}c--e, with the corresponding Cr-$d$-projected FS presented in Supplementary Fig.~S21. The calculations are performed at the DFT, DFT+SOC, and DFT+$U$+SOC levels of theory, revealing the distinct roles of SOC and electronic correlations. At the DFT level (Fig.~\ref{fig:figure4}c), the Cr-derived FS comprises a large sixfold-warped hole pocket centered at $\Gamma$, together with smaller electron pockets at $K$ and along the Brillouin-zone boundary, characteristic of a multiband semimetal with a relatively high carrier density. Upon including SOC (Fig.~\ref{fig:figure4}d), the electron pockets become slightly distorted while remaining largely preserved, indicating that SOC alone has only a weak effect on the FS topology. The low-energy electronic structure remains dominated by Cr-$t_{2g}$ states, with SOC primarily introducing momentum-dependent spin splitting. The inclusion of electronic correlations within the DFT+$U$+SOC framework ($U_{\mathrm{eff}} = 3.2$ eV) leads to a pronounced reconstruction of the low-energy electronic structure and FS topology, as shown in Fig.~\ref{fig:figure4}e. It undergoes a reduction in SW and effective carrier density, accompanied by strong suppression of the electron pockets at $K$ and $M$. The remaining low-energy states are dominated by smaller hole pockets around $\Gamma$, originating primarily from the $d_{z^2}$, $d_{xy}$, and $d_{x^2-y^2}$ orbital characters, together with a strongly modulated flower-like outer contour mainly composed of $d_{xy}$, $d_{xz}$, $d_{yz}$, and $d_{x^2-y^2}$ orbitals.  The correlated DFT+$U$+SOC electronic structure provides a unified microscopic description of the experimentally observed low-temperature insulating transport and topological Hall effect.

\section{Conclusion} 
 In summary, we report a catalyst-free CVD route for phase-pure, single-crystalline 1T-CrS$_2$ with long-term ambient stability, establishing it as a promising candidate for next-generation two-dimensional van der Waals ferromagnets. Magnetization measurements, supported by first-principles calculations, demonstrate robust intrinsic room-temperature ferromagnetism with pronounced out-of-plane magnetic anisotropy and a Curie temperature calculated to be 390 K, identifying 1T-CrS$_2$ as a rare correlated 3$d$-orbital layered ferromagnet operating above room temperature. We unveiled a correlation-driven semimetal--insulator crossover accompanied by persistent negative magnetoresistance and the emergence of a topological Hall effect below 30 K, revealing an unconventional transport regime beyond conventional ferromagnetism. First-principles DFT+$U$+SOC calculations show that spin--orbit coupling gaps the Dirac-like band crossings, while electronic correlations reconstruct the Fermi surface by suppressing the electron pockets and generating a low-carrier-density state centred around $\Gamma$. The resulting momentum-dependent out-of-plane spin polarization, together with strong Heisenberg exchange interactions and finite magnetic anisotropy, provides a unified microscopic framework for the observed magnetic properties and topological transport.  These findings establish 1T-CrS$_2$ as a unique platform where correlated electronic states, room-temperature ferromagnetism, and emergent topological transport coexist, opening new opportunities for exploring correlation-driven spintronic and topological functionalities in two-dimensional quantum materials.

\section{Methods}
\subsection{Experimental Methods}
Phase-pure 1T-CrS$_2$ crystals were synthesized by catalyst-free CVD under ambient pressure in a horizontal tube-in-tube furnace. Chromium chloride (99\%, Sigma-Aldrich) and sulfur (99.98\%, Sigma-Aldrich) powders were used as the Cr and S precursors, respectively, and placed in separate heating zones $\approx$25~cm apart. The precursors were evaporated from independently controlled heating zones and transported by a flowing 95\% Ar with 5\% H$_2$ carrier gas onto a face-down c-plane sapphire, Al$_2$O$_3$ (0001) substrate at 900$^\circ$C, while the sulfur source was maintained at 180$^\circ$C. The resulting CrS$_2$ crystals exhibited hexagonal, truncated-triangular, and triangular-shaped with lateral dimensions exceeding 15~$\mu$m (Supplementary Fig.~S20).

The crystal structure of the as-grown CrS$_2$ crystals was characterized by XRD (Rigaku SmartLab) using Cu K$\alpha$ radiation ($\lambda = 1.5406$~\AA). HR-TEM and SAED were performed on an aberration-corrected microscope operated at 200~kV to verify the 1T crystal structure and phase purity. Optical microscopy and AFM (Bruker) were used to characterize the surface morphology and thickness of the as-grown and exfoliated 1T-CrS$_2$ flakes. AFM measurements were performed in tapping mode. Raman spectra (Horiba LabRAM HR) were acquired with a 532~nm excitation laser at room temperature. The laser power was maintained below 0.5~mW to minimize laser-induced heating. The as-grown CrS$_2$ flakes were transferred onto Si/SiO$_2$ substrates or TEM grids using a PMMA-assisted wet-transfer process. A 495K PMMA layer was spin-coated onto the sample and baked before releasing the PMMA/CrS$_2$ stack by etching the SiO$_2$ layer in buffered HF for $\sim$25~min. The released film was rinsed twice in deionized water and transferred onto RCA-cleaned Si/SiO$_2$ substrates or TEM grids, followed by annealing at 60$^\circ$C for 15--20~min. The PMMA layer was removed in acetone, followed by rinsing in dichloromethane (CH$_2$Cl$_2$) and isopropyl alcohol (IPA), and the sample was dried under N$_2$ flow. Metallic contacts were fabricated by electron-beam lithography followed by Ti (10~nm)/Au (60~nm) deposition. XPS (ULVAC PHI) measurements were performed with a monochromatic Al K$\alpha$ source ($h\nu$ = 1486.6~eV). The X-ray spot size was varied from 10 to 300~$\mu$m, with an effective probing depth of $\sim$5~nm. Binding energies were calibrated using the adventitious C~1s peak at 284.5~eV. Magnetic and electrical transport properties were measured using a physical property measurement system (PPMS). Longitudinal resistance ($R_{xx}$) and transverse Hall resistance ($R_{xy}$) were simultaneously recorded using a four-probe configuration. 

\subsection{Computational Methods}

The crystal structure of 1T-CrS$_2$ was initially optimized using density functional theory within the projector augmented-wave (PAW) formalism as implemented in the Vienna \textit{ab initio} simulation package (VASP)~\cite{kresse1996efficiency}. The spin-polarized calculations were performed within the generalized gradient approximation in the Perdew--Burke--Ernzerhof (PBE) exchange--correlation functional~\cite{PhysRevLett.77.3865}, with vdWs interactions accounted for using the DFT-D3 scheme of Grimme~\cite{grimme2010consistent}. The Brillouin zone was sampled using a $\Gamma$-centered Monkhorst--Pack $k$-point mesh of $18 \times 18 \times 12$, ensuring accurate integration over reciprocal space. A plane-wave energy cutoff of 520~eV was employed to achieve convergence of the total energy and electronic structure. Structural relaxations were performed by fully optimizing the atomic positions, lattice parameters, and cell volume using a conjugate-gradient algorithm. The electronic self-consistency loop was converged to an energy tolerance of $10^{-6}$~eV, while ionic relaxations were continued until the residual Hellmann--Feynman forces on each atom were reduced below $0.01$~eV/\AA. The optimized crystal structure was subsequently used as the input for further electronic structure and magnetic interaction calculations using the full-potential linear muffin-tin orbital (FP-LMTO) method as implemented in the RSPt code~\cite{wills2000full}. The exchange--correlation functional was treated within the Perdew--Burke--Ernzerhof (PBE). SOC was included using a fully relativistic treatment. The Brillouin zone was sampled using a dense $k$-point mesh to ensure convergence of total energies and magnetic properties. The magnetic exchange interactions were calculated using the Liechtenstein--Katsnelson--Antropov--Gubanov (LKAG) formalism~\cite{LIECHTENSTEIN198765,RevModPhys.95.035004}, as implemented in the RSPt code~\cite{wills2000full}. All exchange interactions within a radial cutoff of 7.0a from the central Cr site were included~\cite{entel2012basic}. Finite-temperature magnetic properties were calculated using Metropolis Monte Carlo simulations implemented in the UppASD package~\cite{Eriksson2017}. Atomistic spin-dynamics simulations were performed on a $32 \times 32 \times 32$ supercell to minimize finite-size effects and ensure statistical convergence. For each temperature, an initial $5\times 10^4$ MC steps were used for equilibration, followed by a subsequent $5\times 10^4$ steps for sampling thermodynamic observables. The Curie temperature was determined from the temperature dependence of the magnetic order parameter using a cooling/heating protocol with 2~K temperature increments.

\subsection*{Acknowledgments}
S.M.O. and M.K. acknowledge support by the Croatian Government and the EU through the European Regional Development Fund under the Competitiveness and Cohesion Operational Programme for projects "Centre for Advanced Laser Techniques" (Grant No. KK.01.1.1.05.0001), and "Materials for clean energy, advanced sensors and quantum technologies" (Grant No. PK.1.1.10.0002). C.T. and A.M. acknowledge the bilateral Croatian-Austrian project funded by Croatian Ministry of Science and Education and the Centre for International Cooperation and Mobility (ICM) of the Austrian Agency for International Cooperation in Education and Research (OeAD-GmbH) under project HR 02/2020. A.N.P. acknowledges DST-Nano Mission Grant No. DST/NM/TUE/QM10/2019 and the TRC facilities of S. N. Bose National Centre for Basic Sciences. M.N.H. and H.C.H. acknowledge computational resources provided by the National Academic Infrastructure for Supercomputing in Sweden (NAISS) and financial support from the European Union through the MaMMoS project (Grant Agreement No. 101135546).

\subsection*{Competing interests}

The authors declare no competing interests.

\bibliographystyle{apsrev4-2}
\bibliography{references}

\clearpage
\onecolumngrid

\section{Supplementary Information}


\subsection{Calculation for lattice parameters from SAED image of 1T-CrS$_2$}

For 1T-CrS$_2$, the structure is octahedral, which corresponds to a hexagonal lattice. The interplanar spacing relation for a hexagonal system is given by:

\begin{equation}
\frac{1}{d_{hkl}^2} = \frac{4}{3} \frac{(h^2 + hk + k^2)}{a^2} + \frac{l^2}{c^2}
\end{equation}

For the (100) plane:
\[
h = 1, \quad k = 0, \quad l = 0
\]

Thus, the equation simplifies to:
\[
\frac{1}{d_{100}^2} = \frac{4}{3a^2}
\]

Rearranging for lattice constant \(a\):

\begin{equation}
a = \frac{2}{\sqrt{3}} \, d_{100}
\end{equation}

We calculate the $d_{100}$ value from high resolution SAED pattern of 1T CrS$_2$, 
\[
d_{100} = 0.29 \, \text{nm}
\]

Substituting $d_{100} = 0.29$ at equation $(2)$ gives

\begin{align*}
a &= \frac{2}{\sqrt{3}} \times 0.29 \\
  &\sim 1.1547 \times 0.29 \\
  &\sim 0.34 \, \text{nm}
\end{align*}

\textbf{Finally,}
\[
a \sim 0.34 \, \text{nm} \quad (\text{or } 3.4 \, \text{\AA})
\] which very much consistent with our theoritical optimized 1T-CrS$_2$ lattice consatnt,'a' value which is 3.346 \text{\AA}

\begin{figure}[H]
    \centering
    \includegraphics[width=0.8\textwidth]{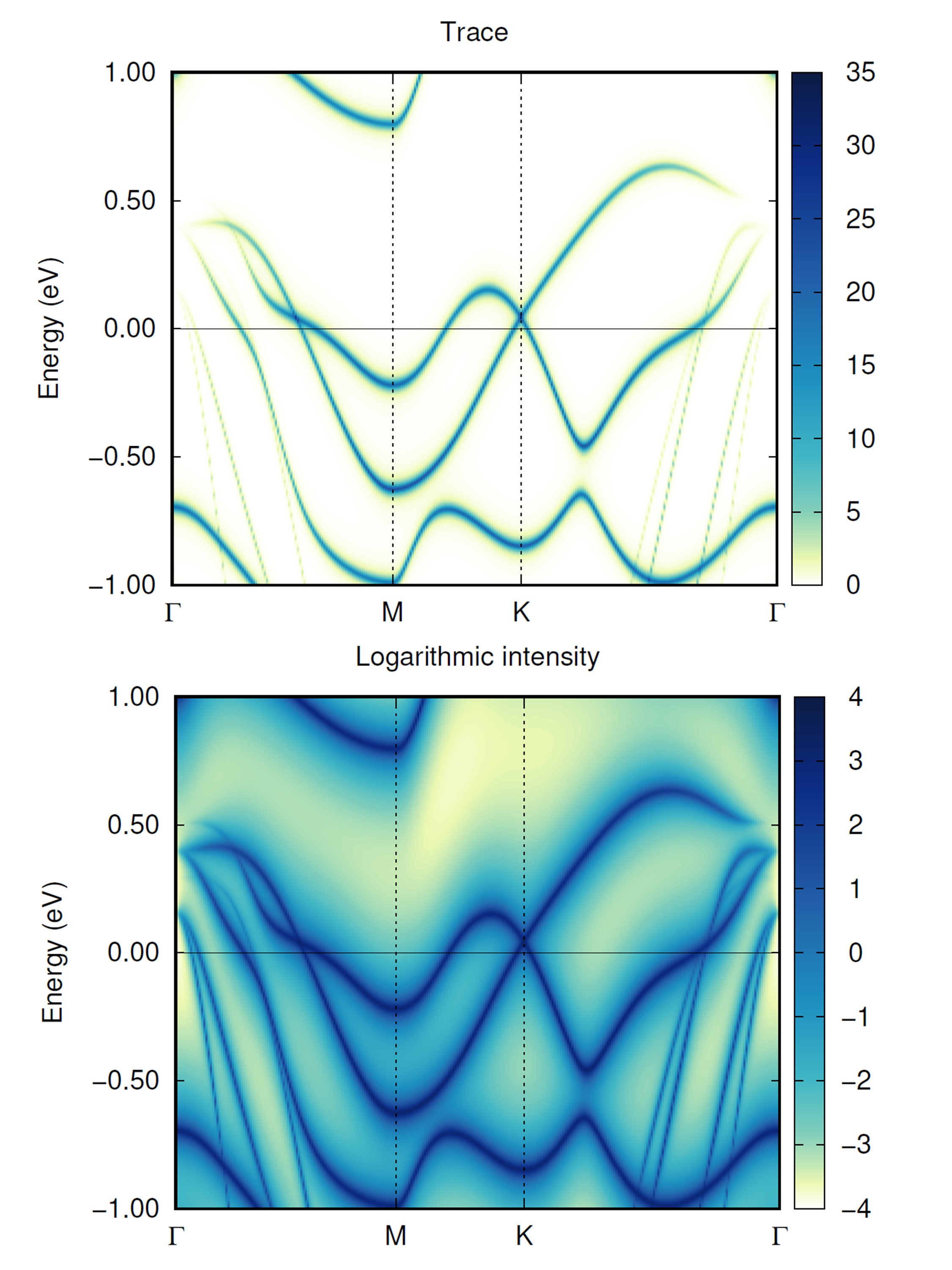}
    \caption{Partial spectral function of Cr 3$d$ states in 1T-CrS$_2$ calculated using DFT without SOC.}
    \label{fig:Figure_S1}
\end{figure}

\begin{figure}[H]
    \centering
    \includegraphics[width=0.8\textwidth]{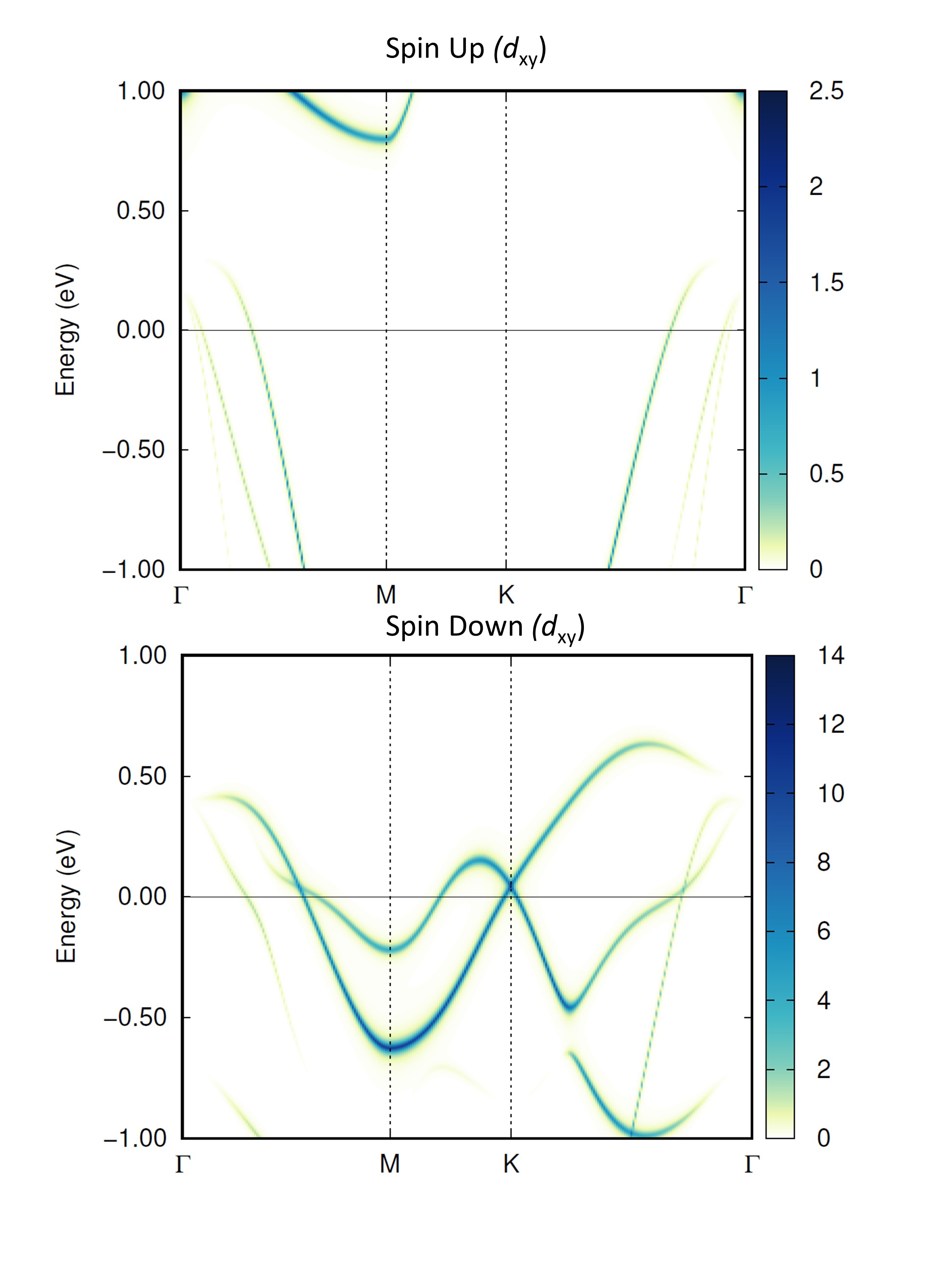}
    \caption{Orbital-resolved spectral function for the Cr $d_{xy}$ orbital of 1T-CrS$_2$, obtained from spin-polarized DFT calculations without SOC.}
    \label{fig:S2}
\end{figure}

\begin{figure}[H]
    \centering
    \includegraphics[width=0.8\textwidth]{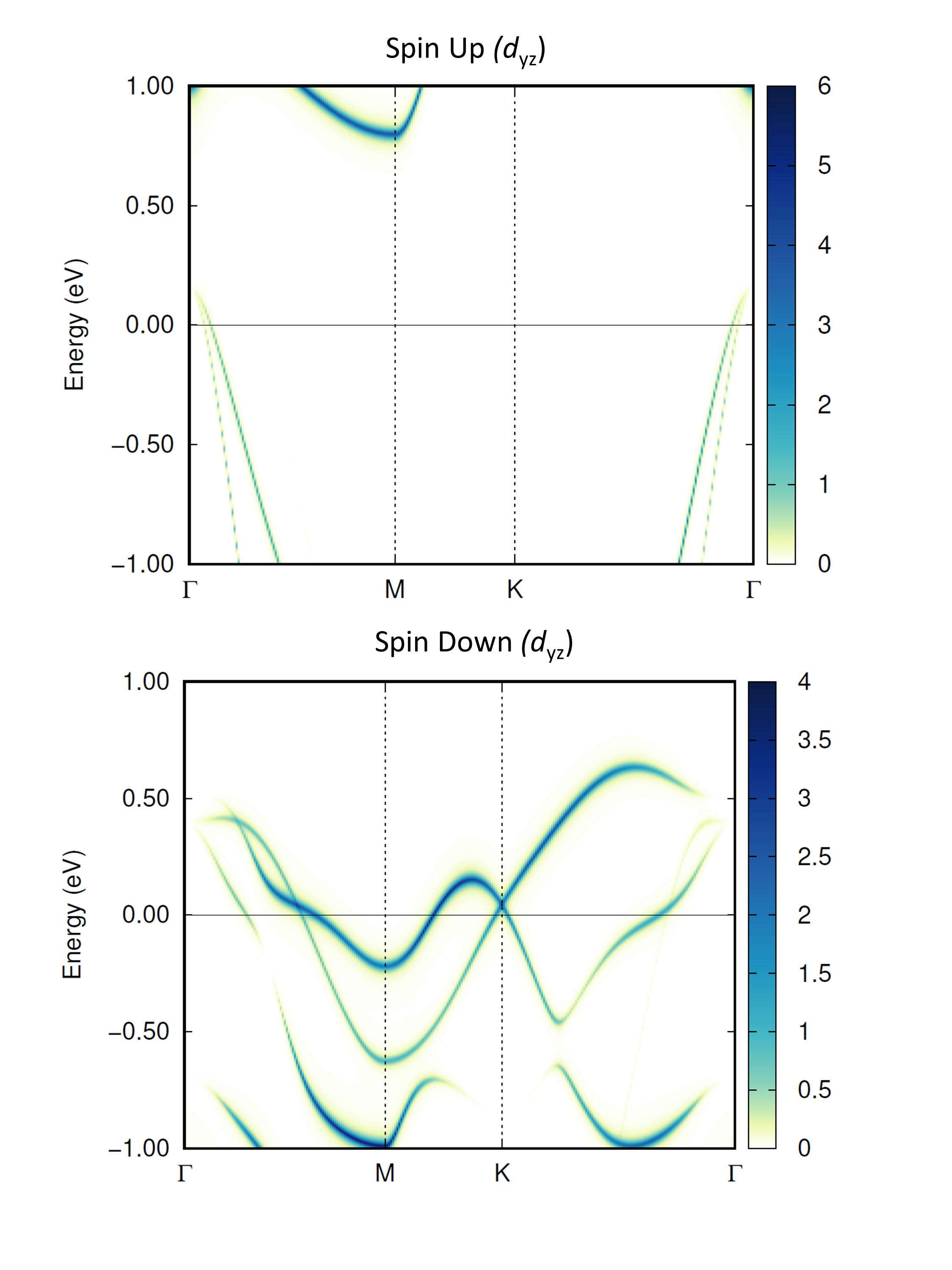}
    \caption{Orbital-resolved spectral function for the Cr $d_{yz}$ orbital of 1T-CrS$_2$, obtained from spin-polarized DFT calculations without SOC.}
    \label{fig:S2}
\end{figure}

\begin{figure}[H]
    \centering
    \includegraphics[width=0.8\textwidth]{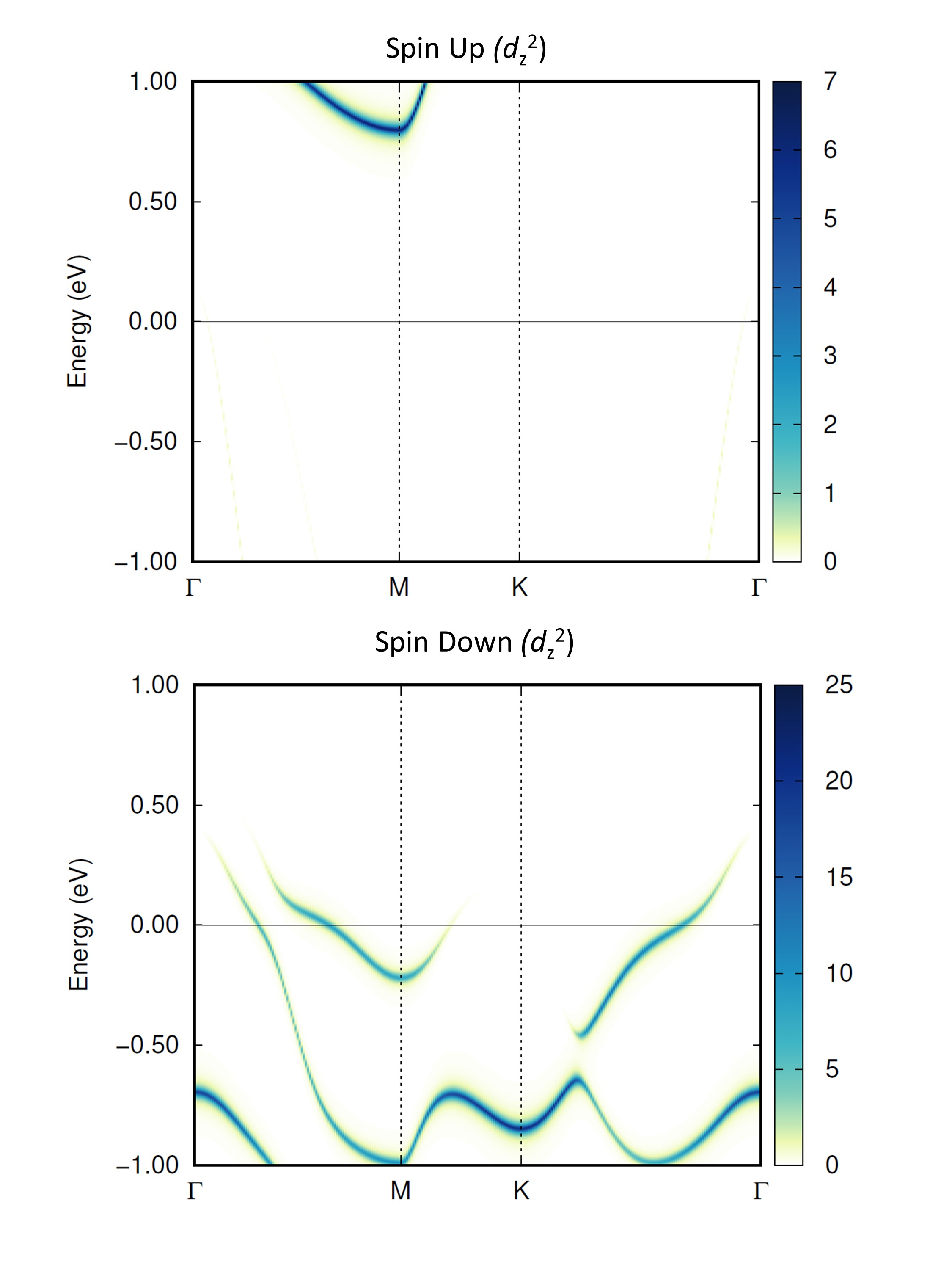}
    \caption{Orbital-resolved spectral function for the Cr $d_{z^2}$ orbital of 1T-CrS$_2$, obtained from spin-polarized DFT calculations without SOC.}
    \label{fig:S2}
\end{figure}

\begin{figure}[H]
    \centering
    \includegraphics[width=0.8\textwidth]{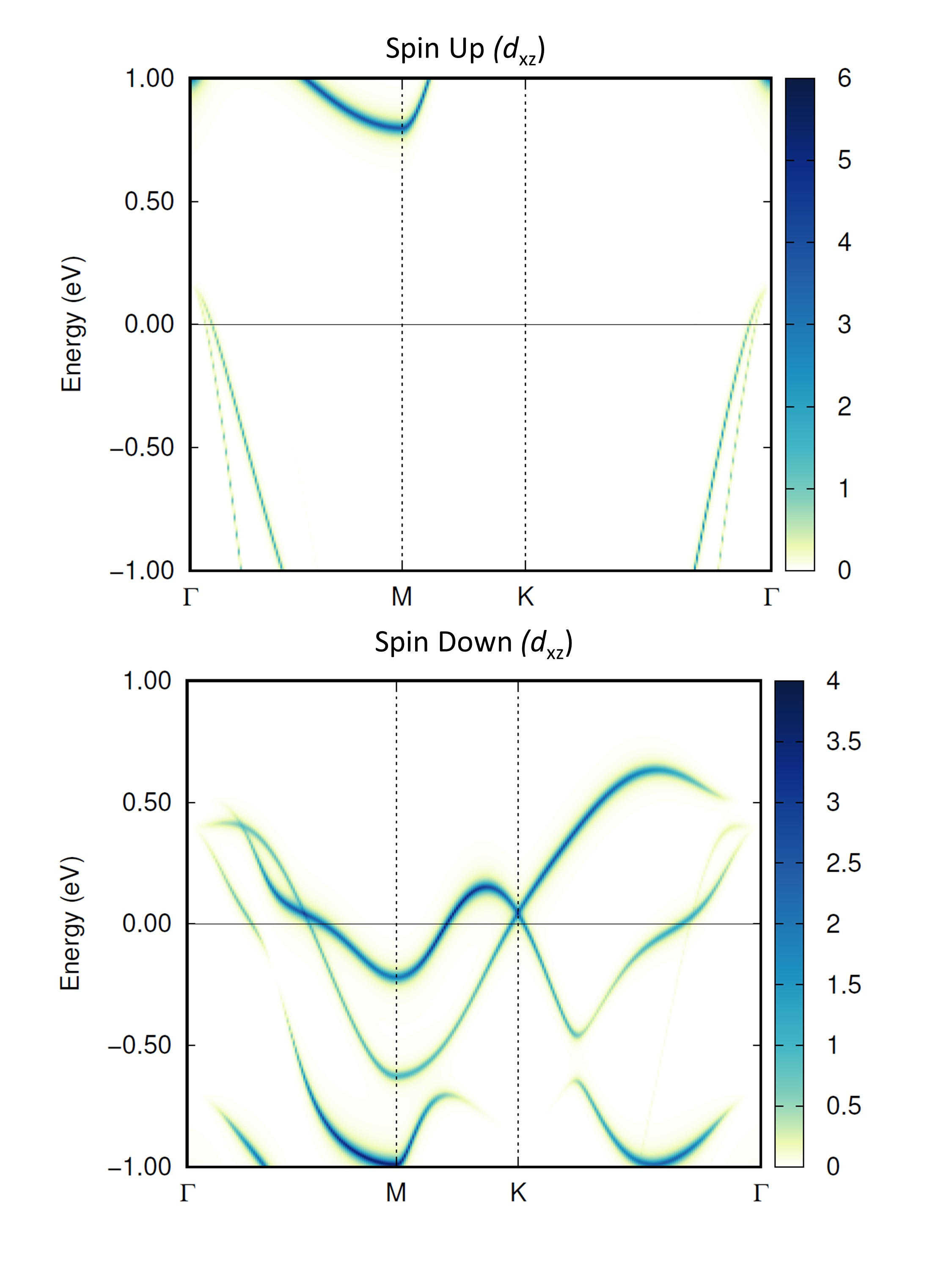}
    \caption{Orbital-resolved spectral function for the Cr $d_{xz}$ orbital of 1T-CrS$_2$, obtained from spin-polarized DFT calculations without SOC.}
    \label{fig:S2}
\end{figure}

\begin{figure}[H]
    \centering
    \includegraphics[width=0.8\textwidth]{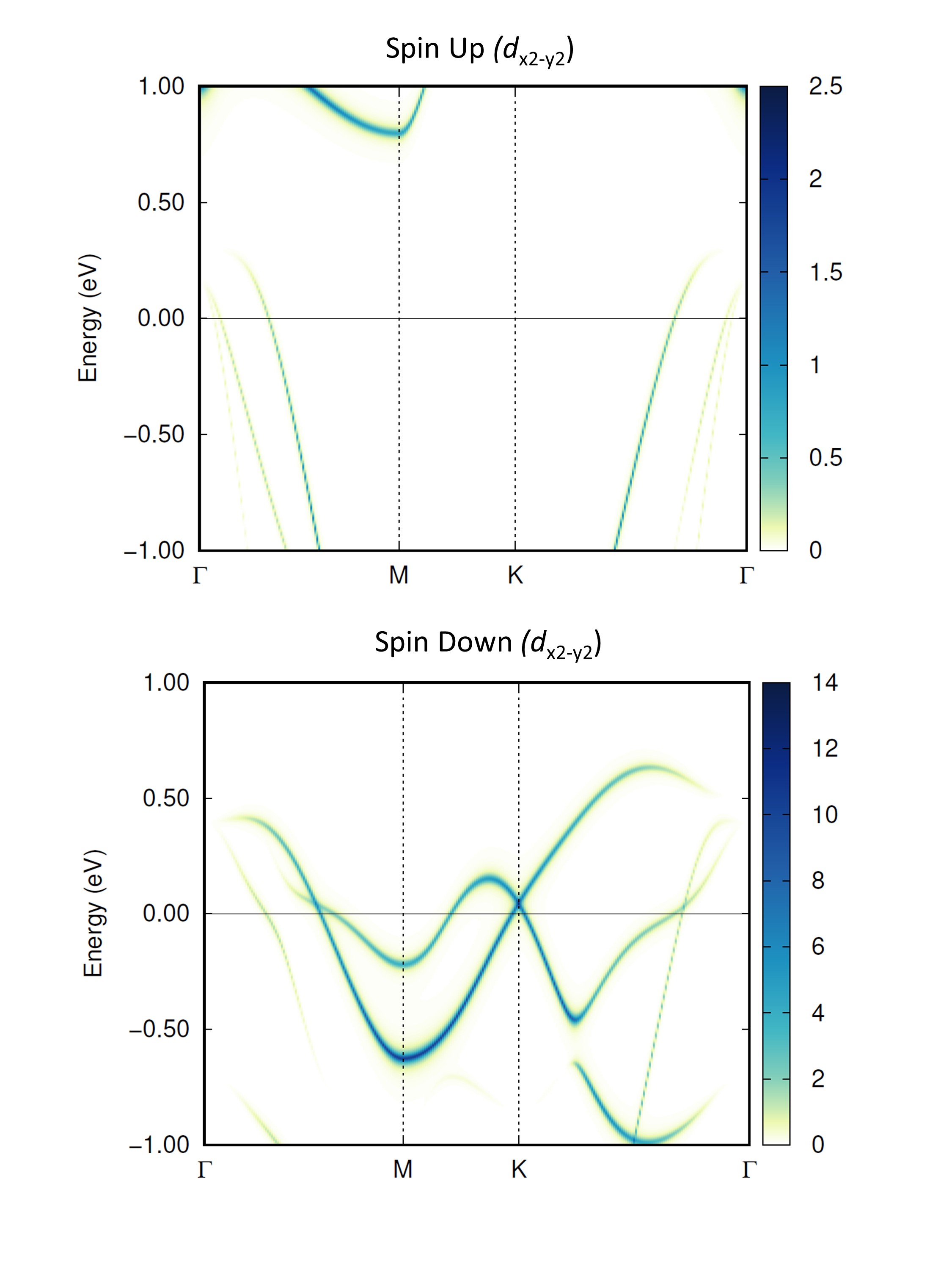}
    \caption{Orbital-resolved spectral function for the Cr $d_{x^2-y^2}$ orbital of 1T-CrS$_2$, obtained from spin-polarized DFT calculations without SOC.}
    \label{fig:S2}
\end{figure}

\begin{figure}[H]
    \centering
    \includegraphics[width=0.8\textwidth]{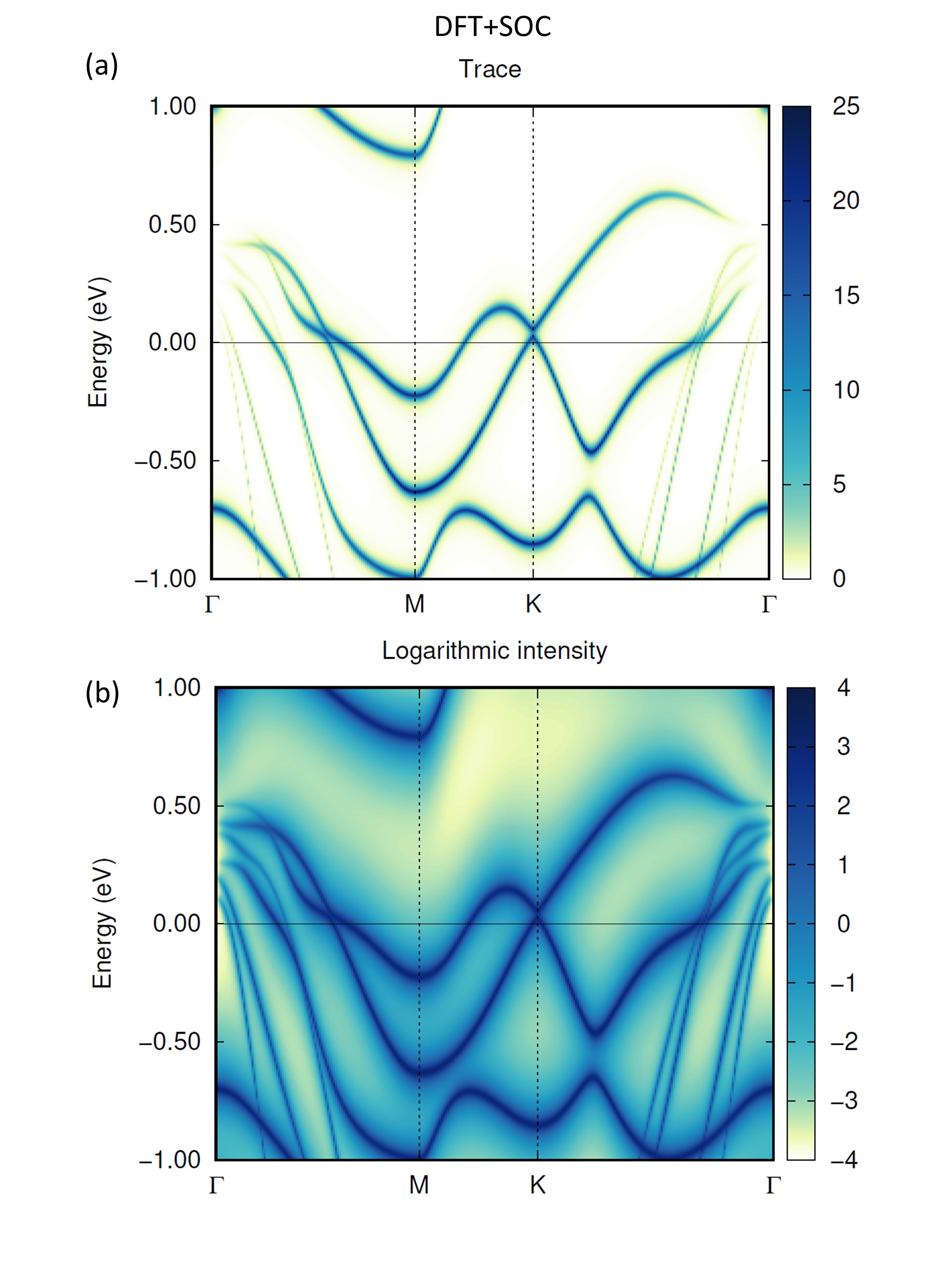}
    \caption{Partial spectral function of Cr 3$d$ states in 1T-CrS$_2$ calculated using }
    \label{fig:Figure_S1}
\end{figure}

\begin{figure}[H]
    \centering
    \includegraphics[width=0.8\textwidth]{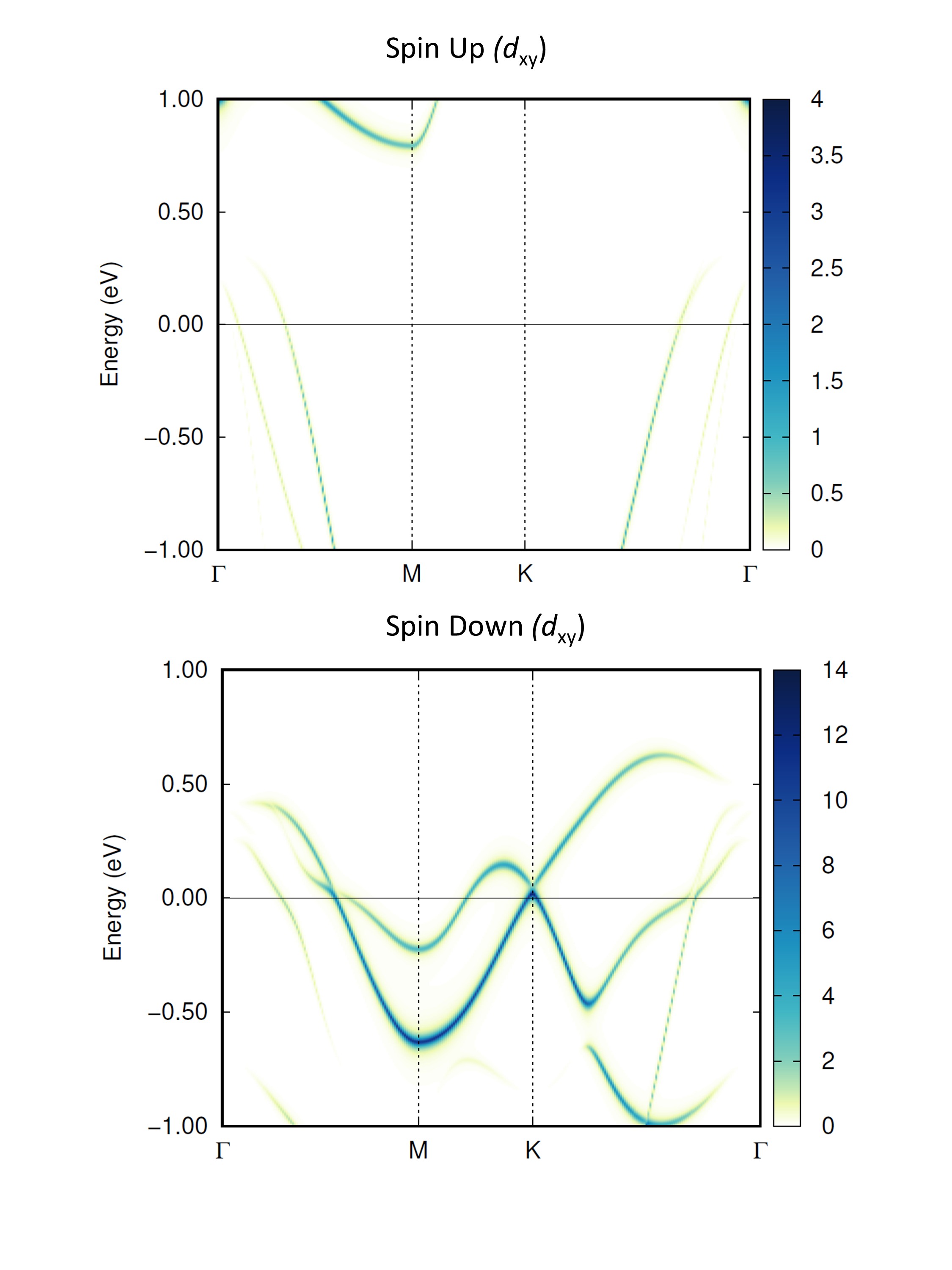}
    \caption{Orbital-resolved DFT+SOC spectral function for the Cr $d_{xy}$ orbital of 1T-CrS$_2$}
    \label{fig:S2}
\end{figure}

\begin{figure}[H]
    \centering
    \includegraphics[width=0.8\textwidth]{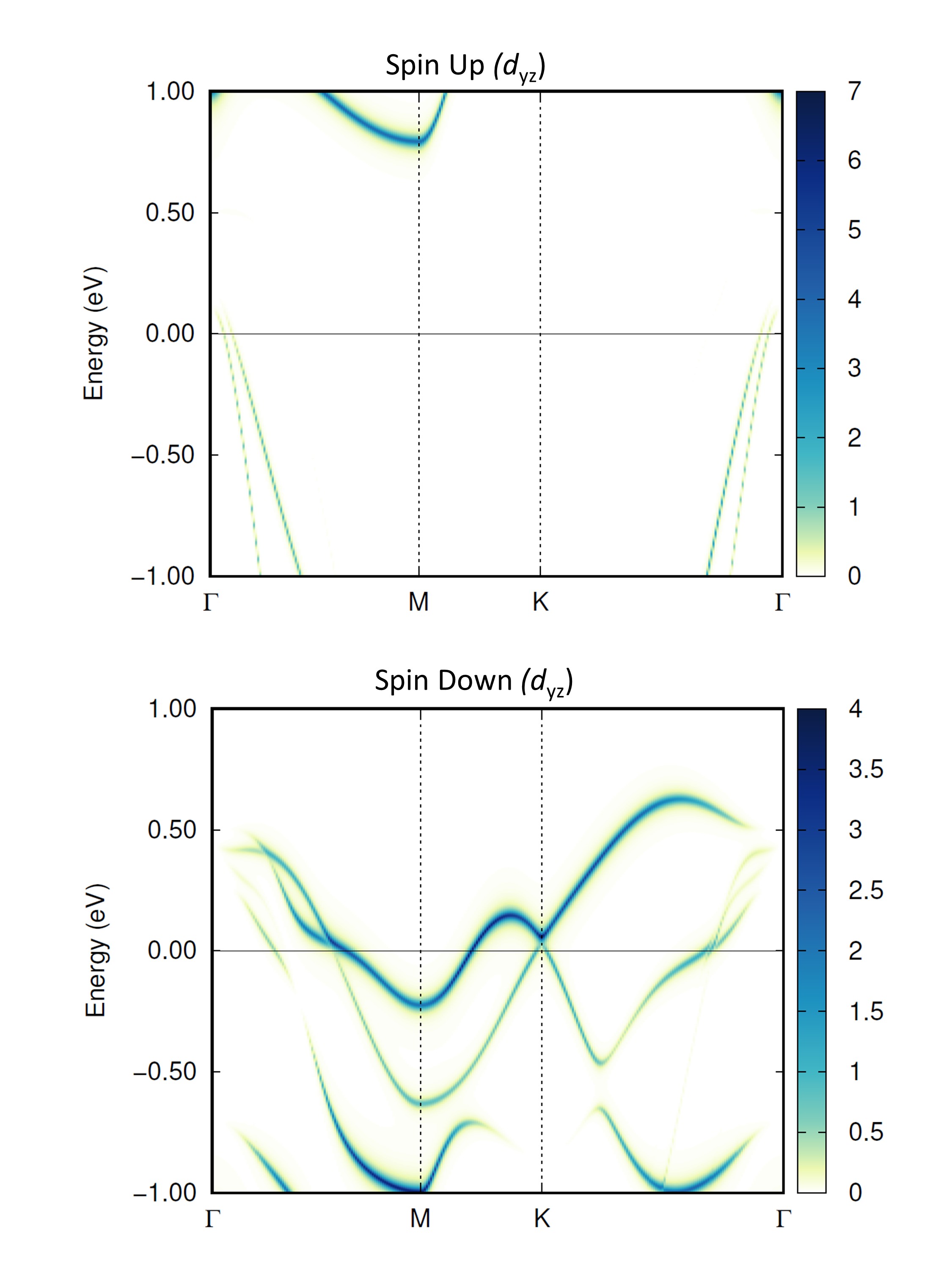}
    \caption{Orbital-resolved DFT+SOC spectral function for the Cr $d_{yz}$ orbital of 1T-CrS$_2$.}
    \label{fig:S2}
\end{figure}

\begin{figure}[H]
    \centering
    \includegraphics[width=0.8\textwidth]{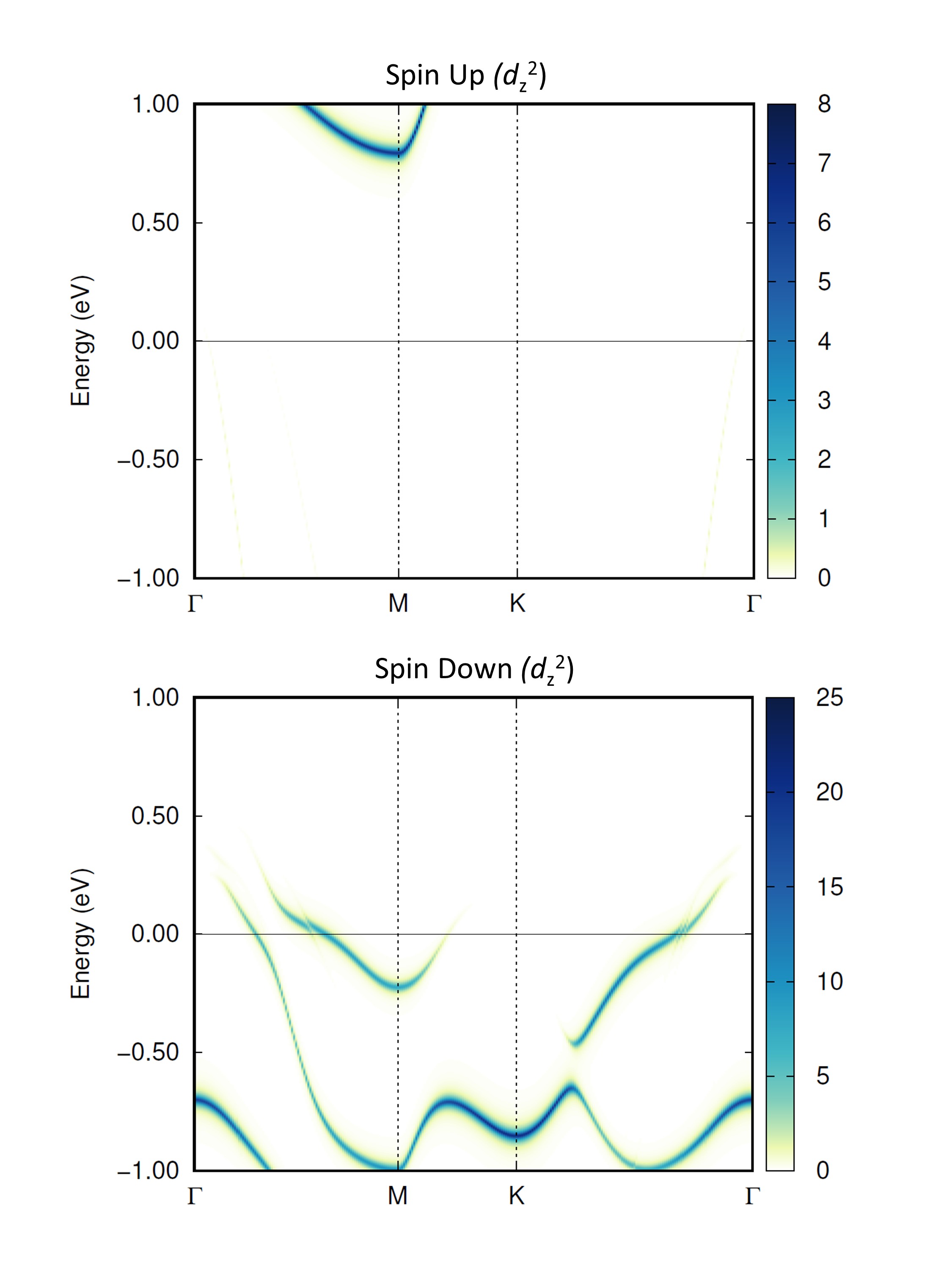}
    \caption{Orbital-resolved DFT+SOC spectral function for the $d_{z^2}$ orbital of 1T-CrS$_2$.}
    \label{fig:S2}
\end{figure}

\begin{figure}[H]
    \centering
    \includegraphics[width=0.8\textwidth]{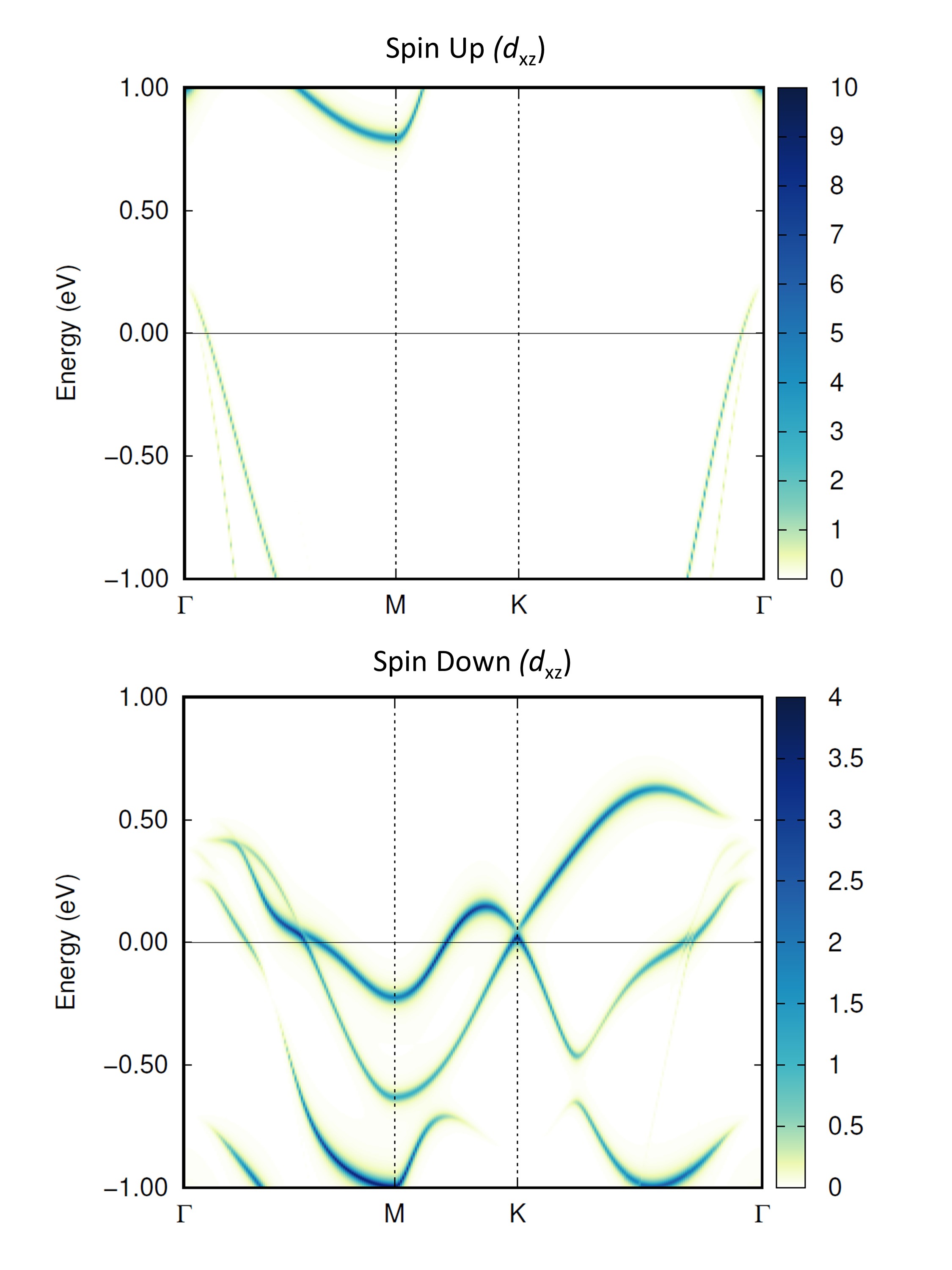}
    \caption{Orbital-resolved DFT+SOC spectral function for the Cr $d_{xz}$ orbital of 1T-CrS$_2$.}
    \label{fig:S2}
\end{figure}

\begin{figure}[H]
    \centering
    \includegraphics[width=0.8\textwidth]{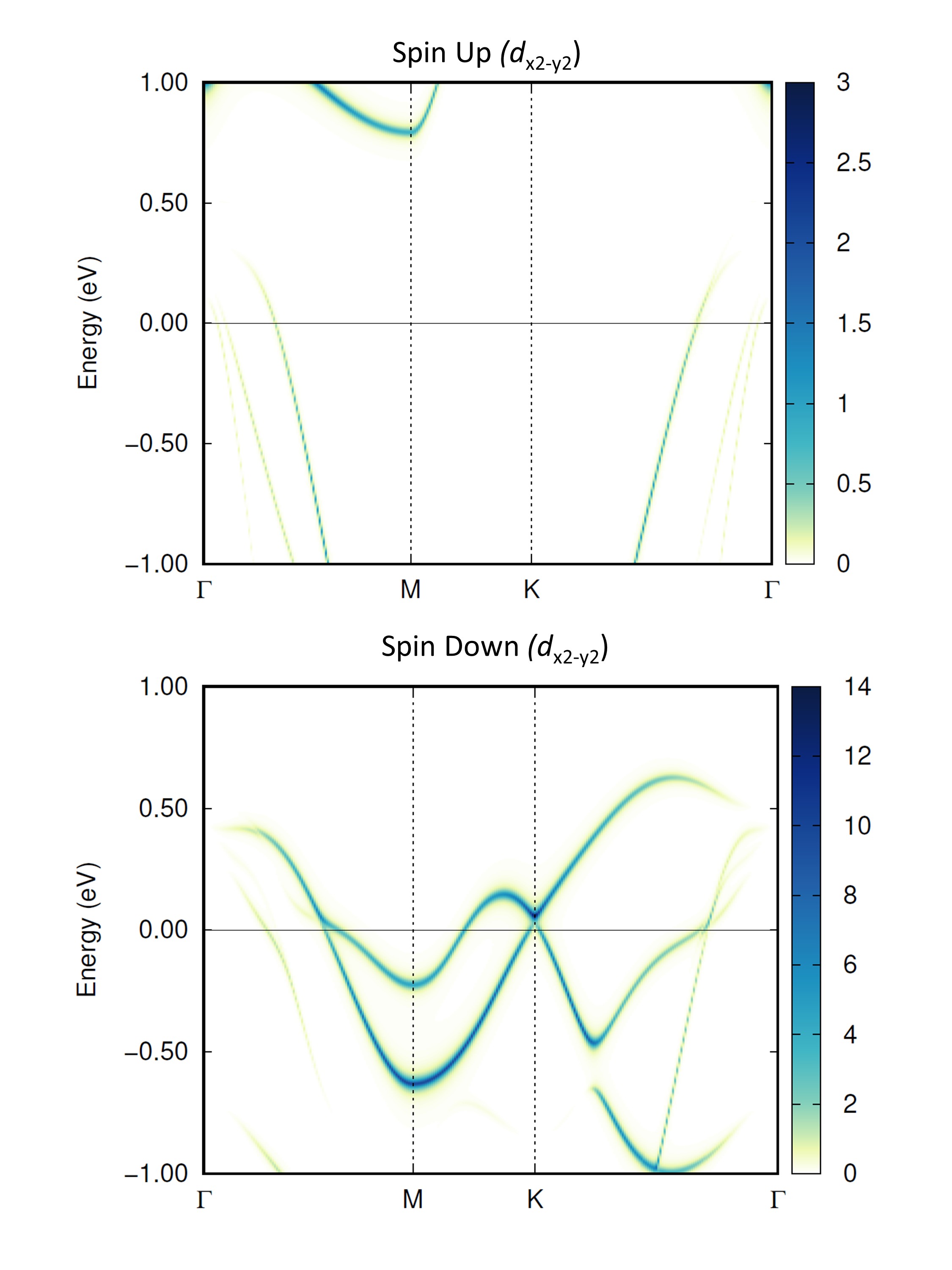}
    \caption{Orbital-resolved DFT+SOC spectral function for the Cr $d_{x^2-y^2}$ orbital of 1T-CrS$_2$}
    \label{fig:S2}
\end{figure}

\begin{figure}[H]
    \centering
    \includegraphics[width=0.8\textwidth]{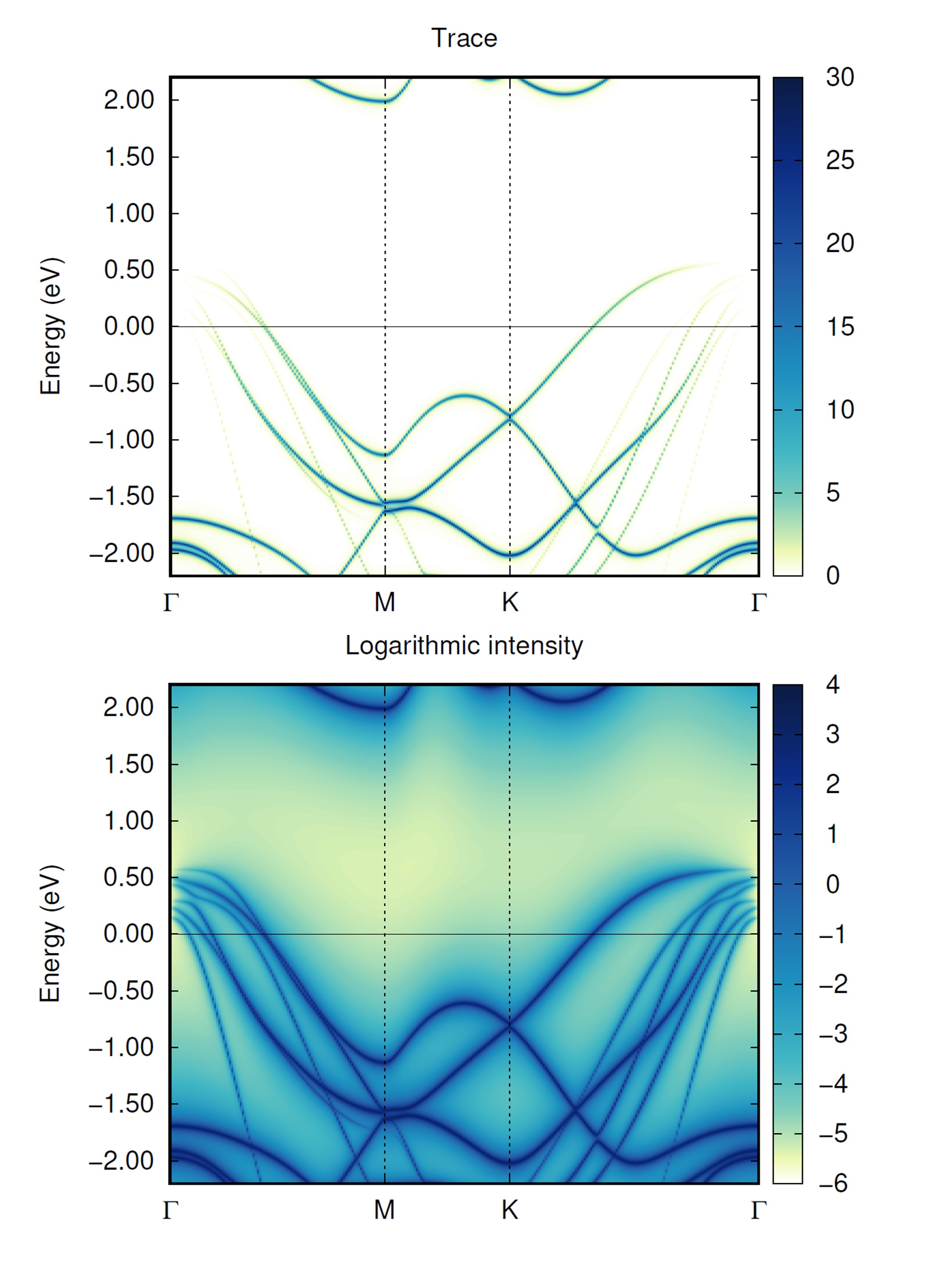}
    \caption{Partial spectral function of Cr 3$d$ states in 1T-CrS$_2$ calculated using DFT+U+SOC method by using full potential RSPt code.}
    \label{fig:S2}
\end{figure}

\begin{figure}[H]
    \centering
    \includegraphics[width=0.8\textwidth]{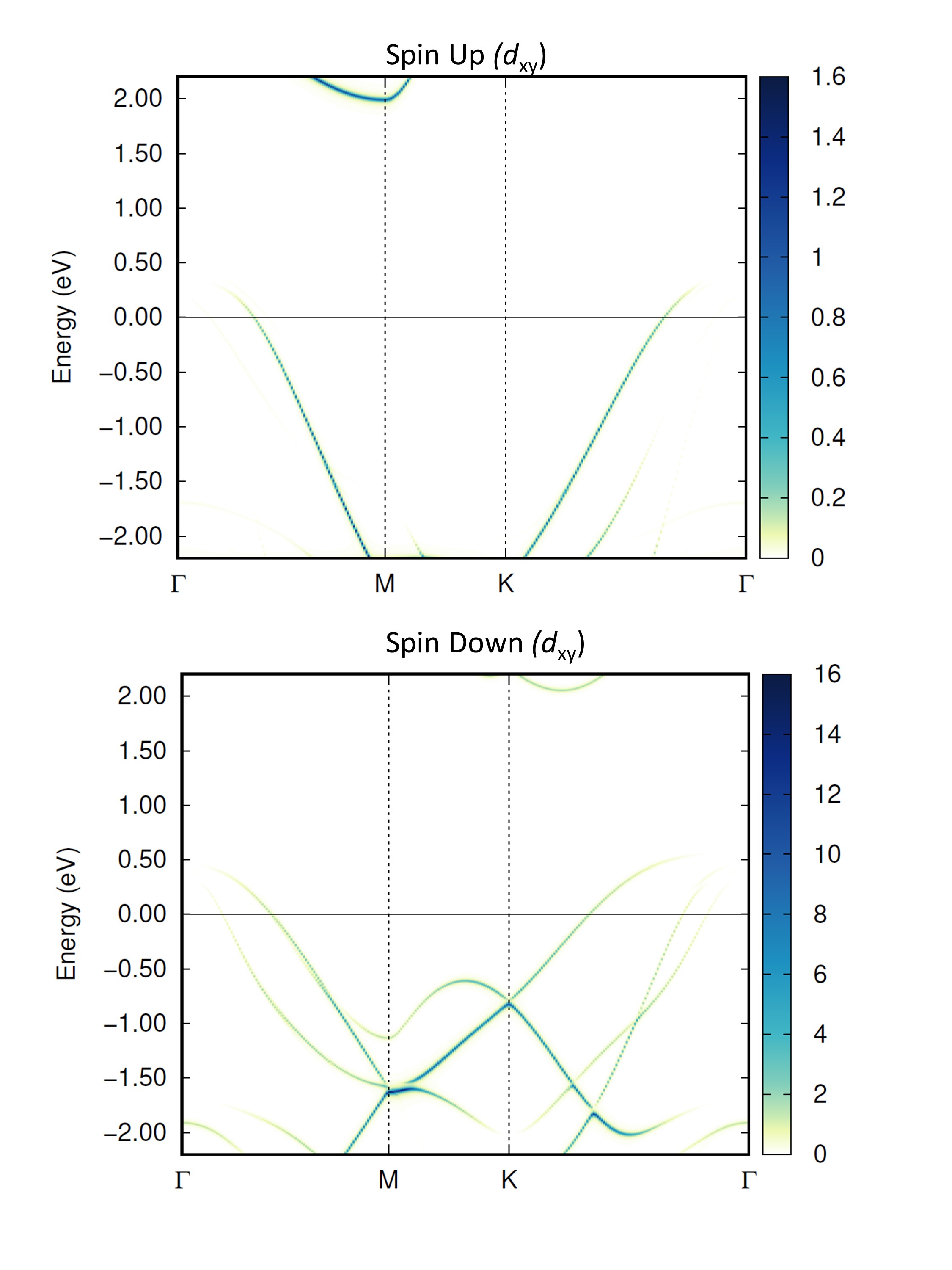}
    \caption{Orbital-resolved DFT+U+SOC spectral function for the Cr $d_{xy}$ orbital of 1T-CrS$_2$.}
    \label{fig:S2}
\end{figure}

\begin{figure}[H]
    \centering
    \includegraphics[width=0.8\textwidth]{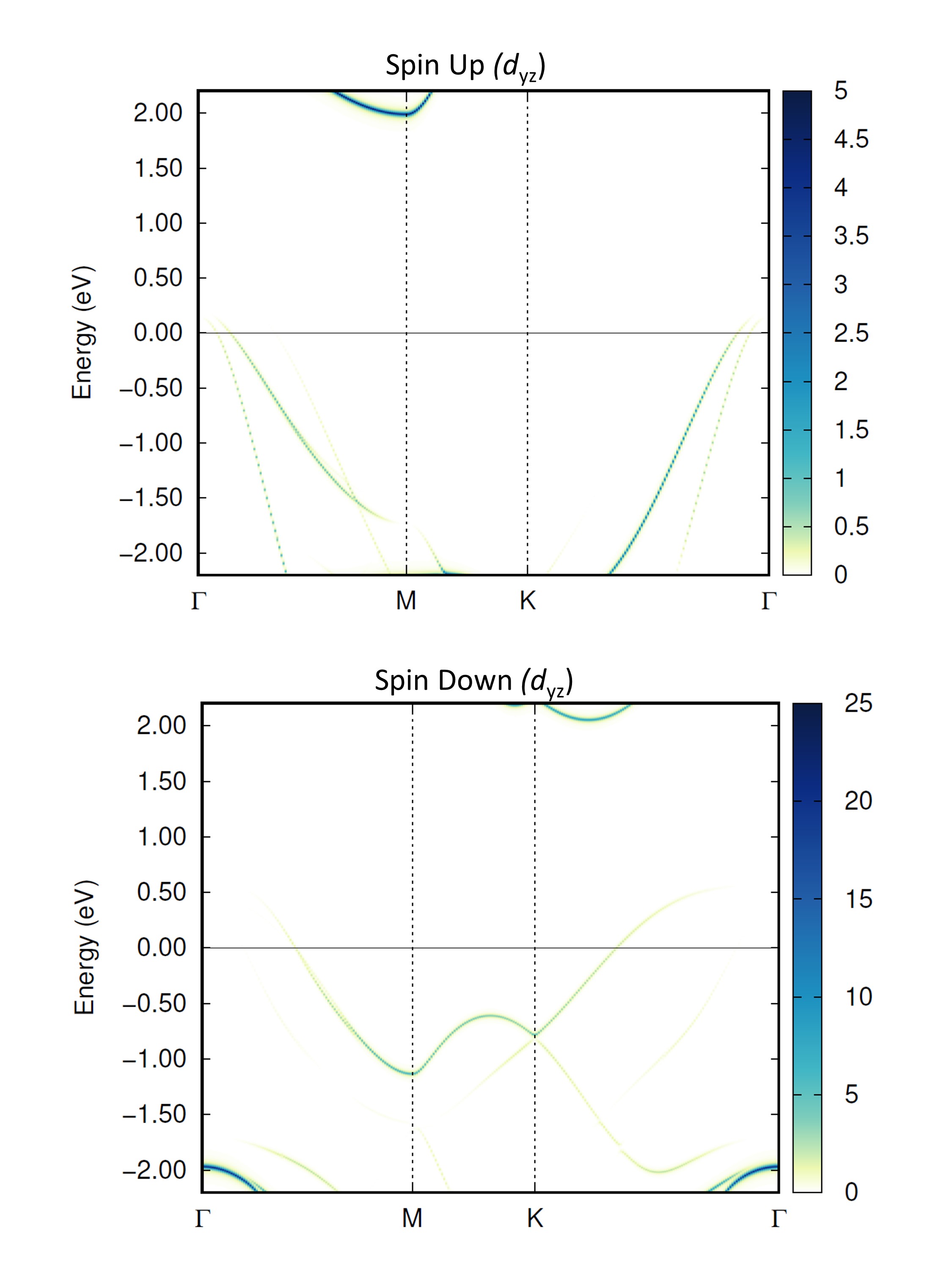}
    \caption{Orbital-resolved DFT+U+SOC spectral function for the Cr $d_{xz}$ orbital of 1T-CrS$_2$.}
    \label{fig:S2}
\end{figure}

\begin{figure}[H]
    \centering
    \includegraphics[width=0.8\textwidth]{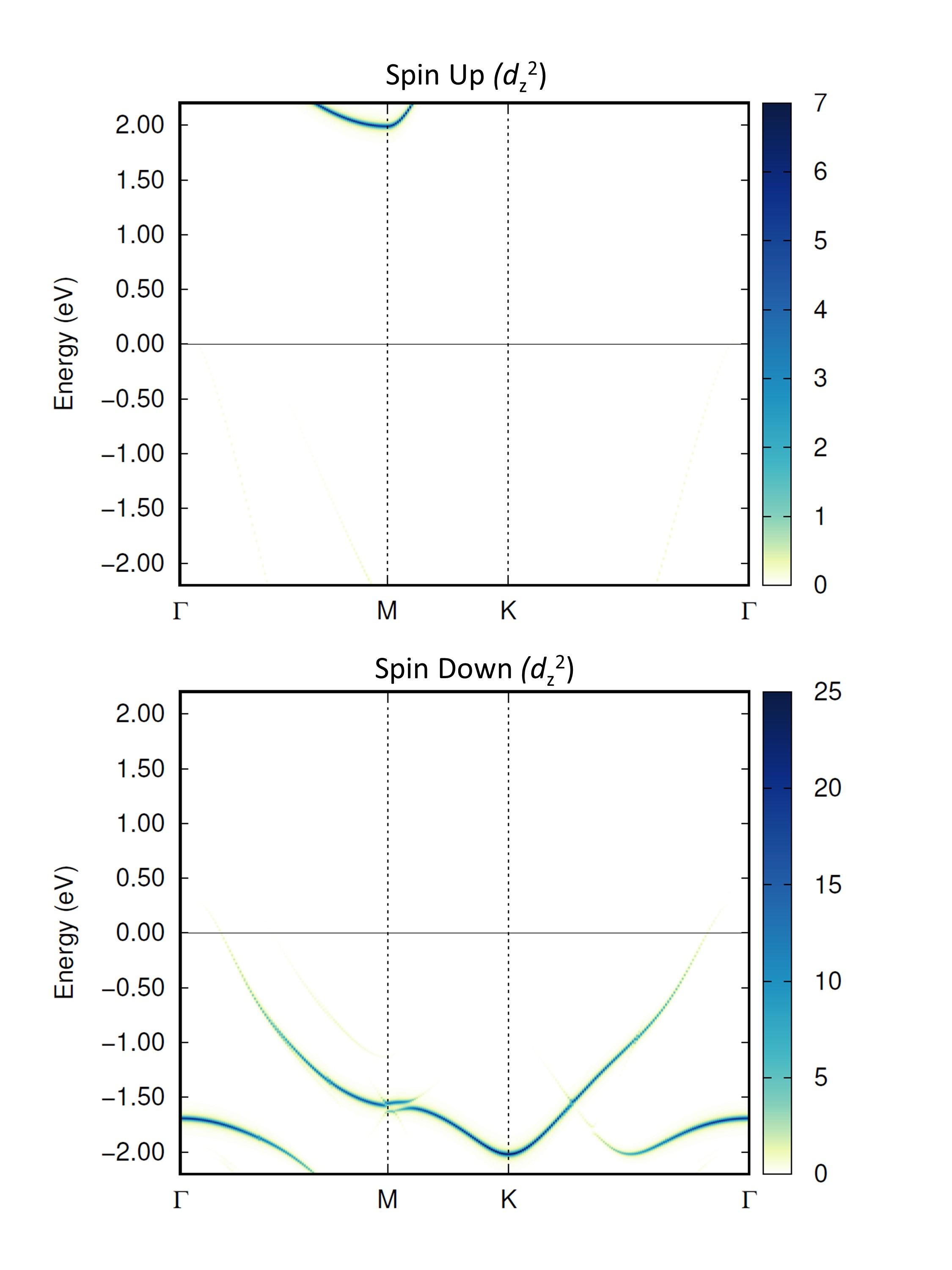}
    \caption{Orbital-resolved DFT+U+SOC spectral function for the Cr $d_{yz}$ orbital of 1T-CrS$_2$.}
    \label{fig:S2}
\end{figure}

\begin{figure}[H]
    \centering
    \includegraphics[width=0.8\textwidth]{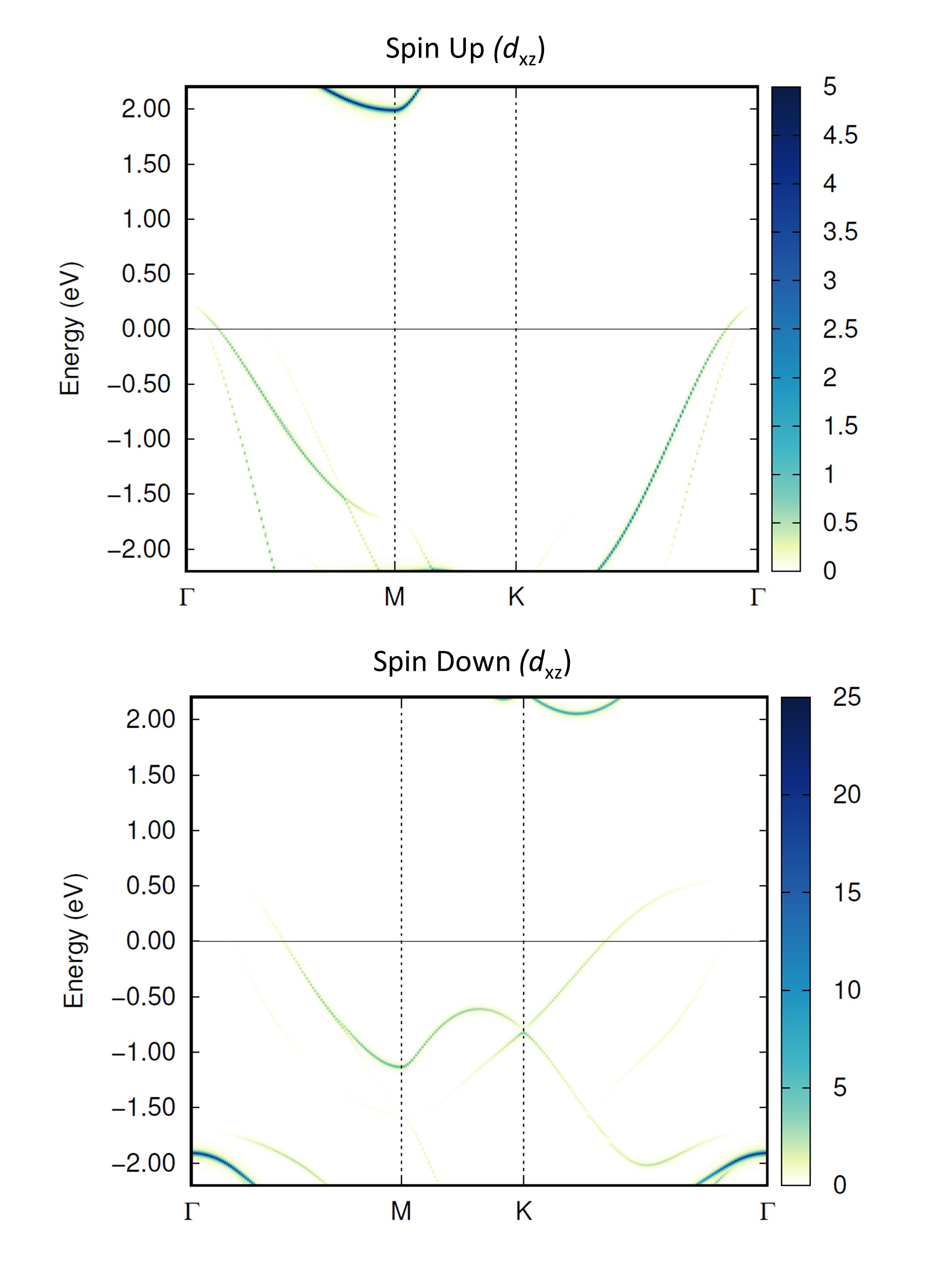}
    \caption{Orbital-resolved DFT+U+SOC spectral function for the Cr $d_{z^2}$ orbital of 1T-CrS$_2$.}
    \label{fig:S2}
\end{figure}

\begin{figure}[H]
    \centering
    \includegraphics[width=0.8\textwidth]{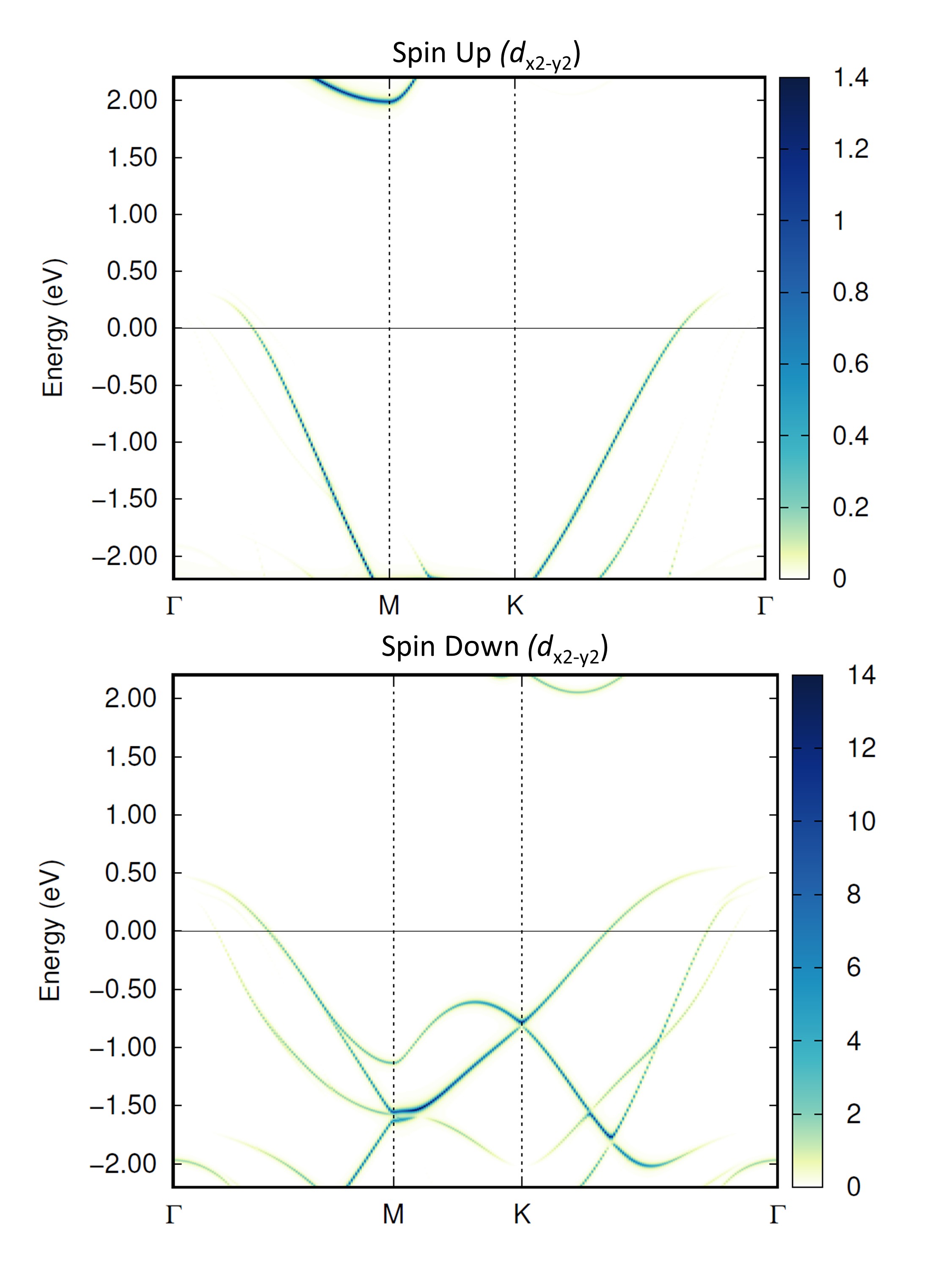}
    \caption{Orbital-resolved DFT+U+SOC spectral function for the Cr $d_{x^2-y^2}$ orbital of 1T-CrS$_2$.}
    \label{fig:S2}
\end{figure}

\begin{figure}[H]
    \centering
    \includegraphics[width=\textwidth]{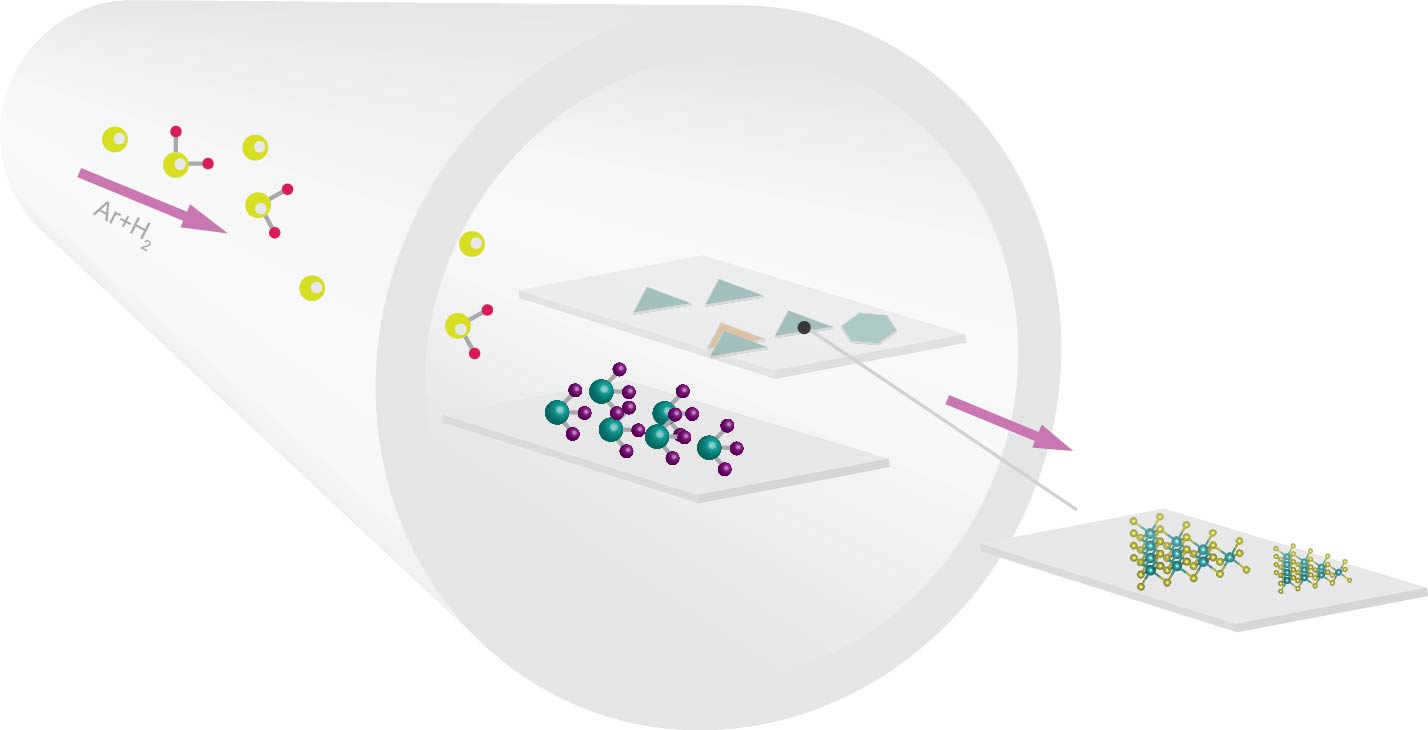}
    \caption{Schematic illustration of the chemical vapor deposition (CVD) growth process of layered 1T-CrS$_2$. Sulfur-containing precursor species are transported by an Ar/H$_2$ carrier gas into the reaction zone, where chromium and sulfur precursors react on the substrate surface, leading to the nucleation and growth of atomically layered 1T-CrS$_2$ domains. The schematic highlights precursor transport, substrate-assisted crystal formation, and the resulting 1T-CrS$_2$ crystal structure after growth.}
    \label{fig:S2}
\end{figure}

\begin{figure}[H]
    \centering
    \includegraphics[width=1.0\textwidth]{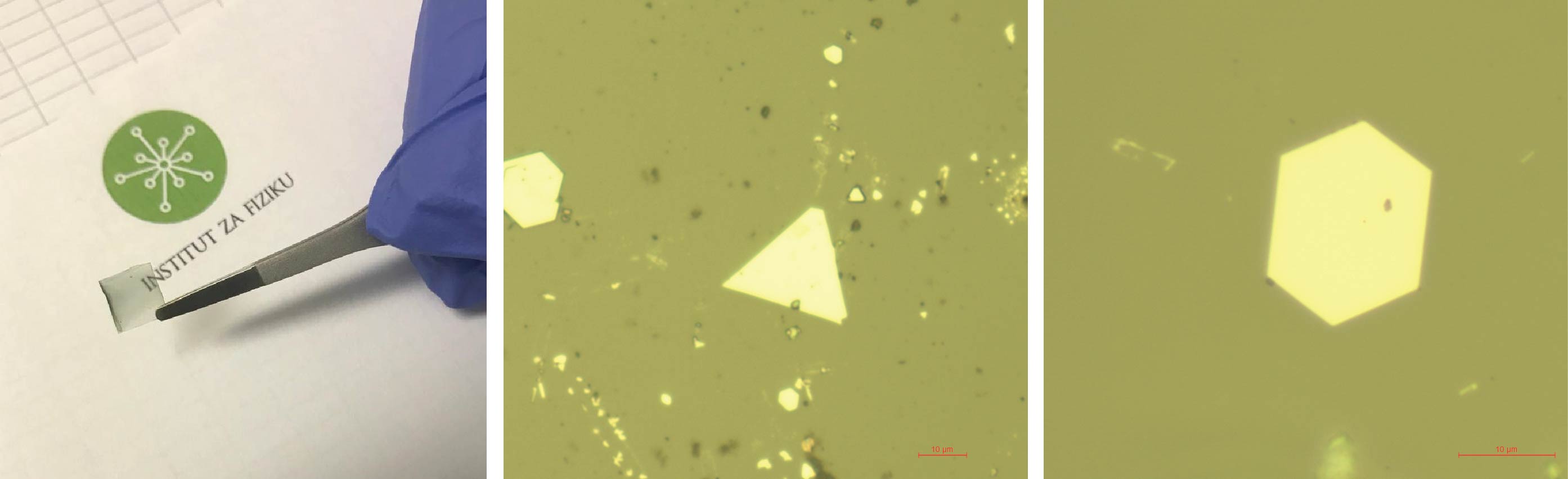}
    \caption{Optical images of as-grown 1T-CrS$_2$ crystals synthesized on transparent sapphire substrates. The left panel shows the transparent sapphire-supported sample after growth, while the middle and right panels display representative triangular and hexagonal single-crystal domains, respectively. The well-defined crystal morphology and sharp edges indicate high crystalline quality, with lateral domain sizes exceeding 15~$\mu$m.}
    \label{fig:S2}
\end{figure}

\begin{figure}[H]
    \centering
    \includegraphics[width=0.5\textwidth]{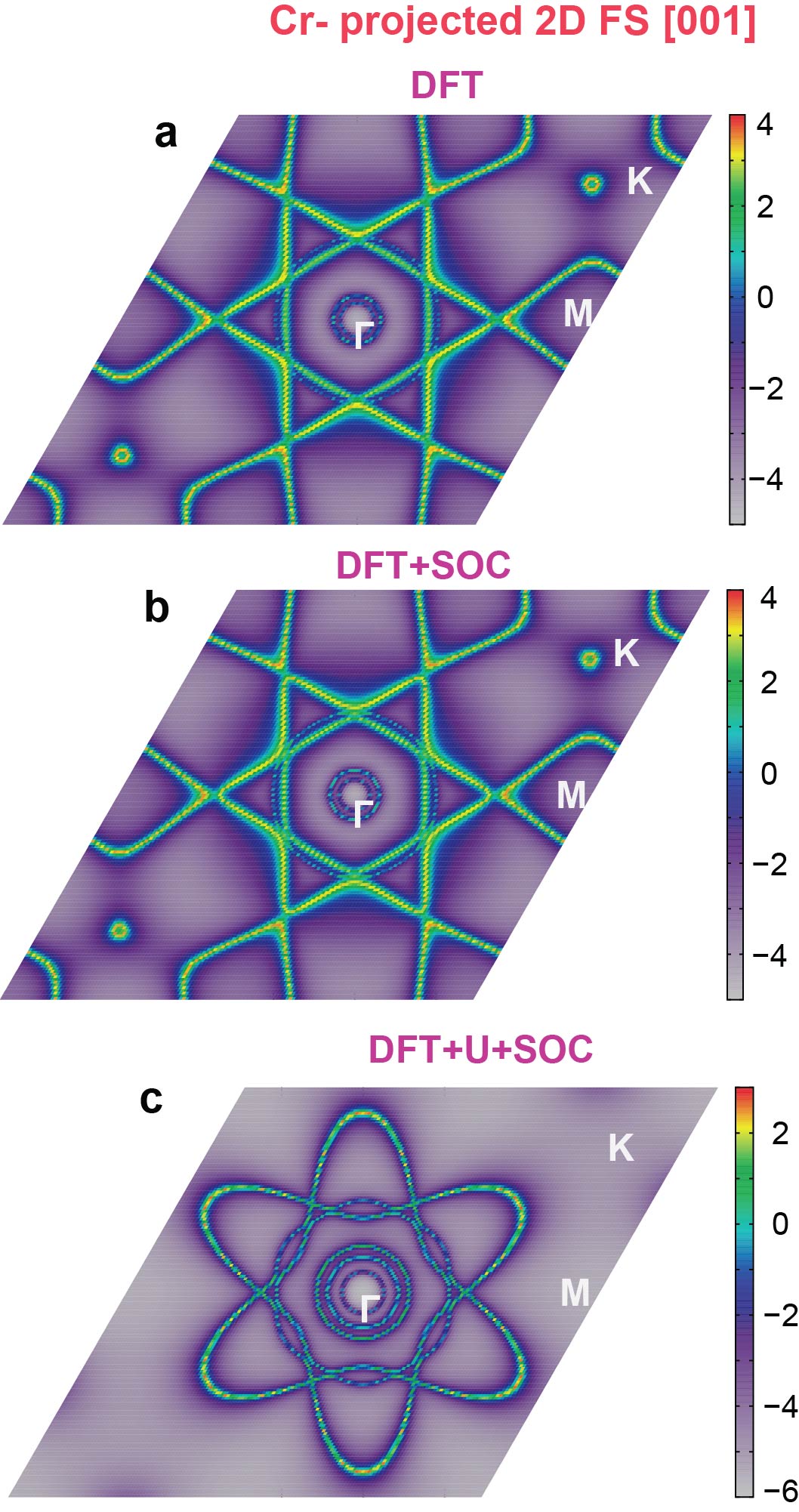}
    \caption{Calculated two-dimensional Cr-projected Fermi surfaces obtained from \textbf{c}, spin-polarized DFT without SOC; \textbf{d}, fully relativistic DFT+SOC; and \textbf{c}, fully relativistic spin-polarized DFT+$U$+SOC calculations.}
    \label{fig:S2}
\end{figure}

\end{document}